%% file: roadmap_main.tex
\documentclass[12pt]{iopart}
\usepackage{amssymb,amsbsy}
\usepackage[pdftex]{graphicx}
\usepackage{hyperref}
\usepackage{url}
\usepackage{color}

\newcommand{\met}{\mbox{g}}
\newcommand{\bmet}{\tilde{\mbox{g}}}

\newcommand{\be}{\begin{equation}}
\newcommand{\ee}{\end{equation}}
\newcommand{\ba}{\begin{eqnarray}}
\newcommand{\ea}{\end{eqnarray}}

\begin{document}

\title[NR/HEP: roadmap for the future]{NR/HEP: roadmap for the future}

\input{AuthorList.tex}



\maketitle

\tableofcontents

\markboth{NR/HEP: roadmap for the future}{NR/HEP: roadmap for the future}

\newpage
\section{Introduction}

Numerical relativity (NR), the gauge/gravity duality and trans-Planckian
scattering, as well as high-energy physics (HEP) in general, have been
tremendously active and successful research areas in recent years.
Strong motivation for the combined study of these fields has arisen
from direct experimental connections: gravitational wave detection,
probing strong interactions at the LHC and RHIC and possibly black hole
production at LHC. Inspired in part by the fairly recent advent of
techniques to evolve BH spacetimes numerically and the unprecedented
opportunities thus opened up to expand and test our understanding of
fundamental physics and the universe, a group of leading experts in
different fields (high energy physics, astrophysics, general relativity,
numerical relativity and phenomenology of gravitational effects on curved
backgrounds) got together from 29 August to 3 September 2011 in Madeira
island (Portugal) to review some of the most exciting recent results and
to discuss important future directions. The meeting was both an excuse
and an opportunity to place at the same table colleagues from formerly
disjoint fields and to discuss the vast amount of possibilities that
exist at the interface.

This overview, written in a ``white-paper'' style, is a summary
of the many interesting discussions at the meeting. Online
presentations can be found at the meeting's website
\textcolor{blue}{\href{http://blackholes.ist.utl.pt/nrhep/}{http://blackholes.ist.utl.pt/nrhep/}}.

We have divided the meeting's discussions into six (not-really-disjoint)
parts, each culminating in a round-table with all participants discussing
the state of the field and visions for the future.  Each round table was
coordinated by one participant who has also been in charge of putting
together a write-up for that particular section. The (somewhat artificial)
division, has resulted in the following topics (round table organizers'
names in parentheses):
numerical methods and results (Lehner),
black hole solutions in generic settings (Reall),
trans-Planckian scattering (Park),
gauge/gravity duality (Chesler),
alternative theories of gravity (Gualtieri) and approximation methods (Sopuerta).  
We feel that the rich content of each of these sections alone is sufficient to impart to the
reader the exciting times that lie ahead.

Finally, we end this short Introduction by thanking the European
Research Council, Funda\c c\~ao Calouste Gulbenkian and FCT--Portugal
for their generous support of this workshop. Special thanks go to Ana
Sousa and Rita Sousa whose help, talent and patience was invaluable,
and to S\'ergio Almeida and Lu\'{i}s Ferreira for the great job done
providing technical and visual assistance.

\noindent Vitor Cardoso

\noindent Leonardo Gualtieri

\noindent Carlos Herdeiro

\noindent Ulrich Sperhake

\newpage

\input NR.tex
\newpage

\input BHs.tex
\newpage

\input Transplanckian_scattering.tex
\newpage

\input Holo.tex

\newpage

\input alternatives.tex
\newpage

\input Approximation.tex

\section*{Acknowledgements}
We warmly thank Marco Sampaio for useful comments and advice.
The support to organize the meeting was generously provided by the
{\it DyBHo--256667} ERC Starting Grant, FCT - Portugal through projects
PTDC/FIS/098025/2008, PTDC/FIS/098032/2008, PTDC/CTE-AST/098034/2008,
CERN/FP/116341/2010 and by Funda\c c\~ao Calouste Gulbenkian.

The authors gratefully acknowledge the support from the Portuguese Science
Foundation (FCT), the Spanish Ministerio de Ciencia e Innovaci\'on (MICINN), the Royal Society,
and the National Science Foundation.
%
R. Emparan and D. Mateos are supported by MEC FPA2010-20807-C02-01 and -02, AGAUR 2009-SGR-168, CPAN  CSD2007-00042 Consolider-Ingenio 2010.
S.B. Giddings  was supported in part by the Department of Energy under Contract DE-FG02-91ER40618 and by grant FQXi-RFP3-1008 from the Foundational Questions Institute (FQXi)/Silicon Valley Community Foundation.
A. Ishibashi was supported by JSPS Grant-in-Aid for Scientific Research (C)No. 22540299.
L.~Lehner thanks CIFAR, NSERC for support through a Discovery Grant and
Perimeter Institute which is supported by the Government of Canada
through Industry Canada and by the Province of Ontario through the
Ministry of Research and Innovation.
C.Lousto gratefully acknowledges the NSF for financial support from Grants
 No. PHY-0722703, PHY-0929114, PHY-0969855, PHY-0903782, and OCI-0832606.
V. Moeller was supported by the Gates Cambridge Trust.
H. Okawa was supported by the Grant-in-Aid for the Global COE Program
``The Next Generation of Physics, Spun from Universality and Emergence''.
P. Pani was supported by the Intra-European Marie Curie contract aStronGR-2011-298297.
S.C. Park is supported by Basic Science Research Program through the National Research Foundation of Korea (NRF) funded by the Ministry of Education, Science and Technology (2011-0010294) and (2011-0029758) and also by Chonnam National University.
M. A. Parker was supported by the Science and Technology Research Council, UK.
H. Reall is a Royal Society University Research Fellow.
C.F. Sopuerta is supported by the Ramon y Cajal Programme of
the Spanish Ministry of Education and Science, the Marie Curie
International Reintegration Grant (MIRG-CT-2007-205005/PHY) within the
7th European Community Framework Programme and Grants AYA-2010-15709
and FIS2011-30145-C03-03 of the Spanish Ministry of Science and Innovation.
N. Yunes was supported by NSF grant PHY-1114374, NASA grant NNX11AI49G, under sub-award 00001944.

\section*{References}
\bibliographystyle{h-physrev4}
\bibliography{References}

\end{document}

%% file: AuthorList.tex
\author{
Vitor Cardoso$^{1,2}$, Leonardo Gualtieri$^3$, Carlos Herdeiro$^4$  and Ulrich Sperhake$^{5,1,6}$ (editors)
}

\author{
Paul M. Chesler$^7$, Luis Lehner$^{8,9}$, Seong Chan Park$^{10}$, Harvey S. Reall$^{11}$, Carlos F. Sopuerta$^{5}$ (Section coordinators)
}

\author{
Daniela Alic$^{12}$,
Oscar J. C. Dias$^{13}$,
Roberto Emparan$^{14,15}$,
Valeria Ferrari$^{3}$,
Steven B. Giddings$^{16}$, 
Mahdi Godazgar$^{11}$, 
Ruth Gregory$^{17}$,  
Veronika E. Hubeny$^{17}$,
Akihiro Ishibashi$^{18,19}$,
Greg Landsberg$^{20}$, 
Carlos O. Lousto$^{21}$,
David Mateos$^{14,15}$, 
Vicki Moeller$^{22}$, 
Hirotada Okawa$^{23}$, 
Paolo Pani$^{1}$, 
M. Andy Parker$^{22}$, 
Frans Pretorius$^{24}$,  
Masaru Shibata$^{23}$,  
Hajime Sotani$^{25}$,  
Toby Wiseman$^{26}$, 
Helvi Witek$^{1}$, 
Nicolas Yunes$^{27,28}$, 
Miguel Zilh\~ao$^{29}$}

\begin{abstract} Physics in curved spacetime describes a multitude of
phenomena, ranging from astrophysics to high energy physics. The last
few years have witnessed further progress on several fronts, including the
accurate numerical evolution of the gravitational field equations, which
now allows highly nonlinear phenomena to be tamed. Numerical relativity
simulations, originally developed to understand strong field astrophysical
processes, could prove extremely useful to understand  high-energy physics
processes like trans-Planckian scattering and gauge-gravity dualities. We
present a concise and comprehensive overview of the state-of-the-art
and important open problems in the field(s), along with guidelines for
the next years. This writeup is a summary of the ``NR/HEP Workshop''
held in Madeira, Portugal from August 31st to September 3rd 2011.
\end{abstract}

\newpage

\address{$^{1}$ Centro Multidisciplinar de Astrof\'\i sica --- CENTRA, Departamento de F\'\i sica, Instituto Superior T\'ecnico --- IST,
Universidade T\'ecnica de Lisboa - UTL, Av. Rovisco Pais 1, 1049-001 Lisboa, Portugal}
\address{$^{2}$ Department of Physics and Astronomy, The University of Mississippi,
  University, MS 38677-1848, USA}
\address{$^{3}$ Dipartimento di Fisica, Universit\`a di Roma ``Sapienza'' \& Sezione,  INFN Roma1, P.A. Moro 5, 00185, Roma, Italy}
\address{$^{4}$ Departamento de F\'\i sica da Universidade de Aveiro and I3N, Campus de Santiago, 3810-183 Aveiro, Portugal}
\address{$^{5}$ Institut de Ci\`encies de l'Espai (CSIC-IEEC), Facultat de Ci\`encies, Campus UAB, Torre C5 parells, Bellaterra, 08193 Barcelona, Spain}
\address{$^{6}$ California Institute of Technology, Pasadena, CA  91125, USA }

\address{$^{7}$ Department of Physics, MIT, Cambridge, MA 02139, USA} 
\address{$^{8}$ Perimeter Institute for Theoretical Physics,Waterloo, Ontario N2L 2Y5, Canada}
\address{$^{9}$ Department of Physics, University of Guelph, Guelph, Ontario N1G 2W1, Canada}
\address{$^{10}$ Department of Physics, Chonnam National University, 300 Yongbong-dong, Buk-gu, Gwangju, 500-757, Korea}
\address{$^{11}$ DAMTP, Centre for Mathematical Sciences, Wilberforce Road, Cambridge CB3 0WA, UK}

\address{$^{12}$ Albert-Einstein-Institut, Max-Planck-Institut f\"ur Gravitationsphysik, D-14476 Golm, Germany}
\address{$^{13}$ Institut de Physique Th\'eorique, CEA Saclay, CNRS URA 2306, F-91191 Gif-sur-Yvette, France} 
\address{$^{14}$ Departament de F\'{\i}sica Fonamental and Institut de Ci\`encies del Cosmos, Universitat de Barcelona,
Mart\'{\i} i Franqu\`es 1, E-08028 Barcelona, Spain} 
\address{$^{15}$ Instituci\'o Catalana de Recerca i Estudis Avan\c cats (ICREA)
Passeig Llu\'{\i}s Companys 23, E-08010 Barcelona, Spain}
\address{$^{16}$ Department of Physics, University of California, Santa Barbara, CA 93106}
\address{$^{17}$ Centre for Particle Theory, Department of Mathematical Sciences
Durham University, South Road, Durham DH1 3LE, U.K.}

\address{$^{18}$ Theory Center, Institute of Particle and Nuclear Studies,
High Energy Accelerator Research Organization (KEK), Tsukuba, 305-0801, Japan}
\address{$^{19}$ Department of Physics, Kinki University, Higashi-Osaka 577-8502, Japan}
\address{$^{20}$ Brown University, Providence, Rhode Island 02912, USA}
\address{$^{21}$ Center for Computational Relativity and Gravitation,
and School of Mathematical Sciences, Rochester Institute of Technology,
85 Lomb Memorial Drive, Rochester, New York 14623}
\address{$^{22}$ Cavendish Laboratory, JJ Thomson Ave, Cambridge CB3 0HE, UK}
\address{$^{23}$ Yukawa Institute for Theoretical Physics, Kyoto University, Kyoto 606-8502, Japan}
\address{$^{24}$ Department of Physics, Princeton University, Princeton, NJ 08544, USA.}
\address{$^{25}$ Division of Theoretical Astronomy, National Astronomical
Observatory of Japan, 2-21-1 Osawa, Mitaka, Tokyo 181-8588, Japan}
\address{$^{26}$ Theoretical Physics Group, Blackett Laboratory, Imperial College, London
SW7~2AZ, U.K.}
\address{$^{27}$ MIT and Kavli Institute, Cambridge, MA 02139, USA}
\address{$^{28}$ Department of Physics, Montana State University, Bozeman, MT 59717, USA}
\address{$^{29}$Centro de F\'\i sica do Porto --- CFP, Departamento de F\'\i sica e Astronomia
Faculdade de Ci\^encias da Universidade do Porto --- FCUP, Rua do Campo Alegre, 4169-007 Porto, Portugal}

%% file: NR.tex
\section{Strong gravity, High Energy Physics and Numerical Relativity} \label{sec:nr}
\begin{center}
Coordinator: Luis Lehner
\end{center}

Extensive experimental and observational programs are underway to test
our understanding of gravity in ``extreme'' regimes. These extreme
regimes are naturally encountered in cosmology, violent astrophysical
events and possibly at high energy particle accelerators if nature is
described by certain higher dimensional scenarios. Furthermore,
the gauge/gravity duality provides an intriguing opportunity to
understand various physical phenomena, naturally described by
field theories, in terms of gravity in Anti-de-Sitter ($AdS$) spacetime
and vice-versa.

These possibilities imply the need to understand gravity in the relevant
regimes, in particular those where approximations (e.g. linearized
studies) fail to describe the behavior of the system.

Doing so requires the ability to obtain appropriate solutions from the
underlying gravitational theory which, in turn, often requires suitable
numerical simulations. Fortunately, in the case of General Relativity in
asymptotically flat, 4-dimensional spacetimes,
this task is largely under control
\cite{Pretorius:2005gq, Campanelli:2005dd,Baker:2005vv}
and will aid in future
astrophysical and cosmological observational prospects. Thus we here
concentrate on summarizing the status and open issues of efforts towards
understanding gravity, at the classical level, in energetic phenomena
related to higher dimensional scenarios. In particular we focus our
attention on efforts to understand gravity --within holographic studies
and TeV-gravity scenarios--, and discuss some important issues, with the
aim to report both on the state-of-the-art and ongoing discussions on
these topics. We restrict our discussion to dynamical systems and issues
related to their study mainly within General Relativity (alternative
gravity theories are discussed in section  \ref{sec:alternative}). As efforts
in this front are just beginning, our discussion cannot possibly be
exhaustive and it is likely that
new issues will arise as progress is made.  Indeed,
drawing from the experience gained in the 4-dimensional setting, it is
safe to expect that the challenges and outcomes
will be as, if not more exciting, than what can be currently anticipated.

\subsection{Framework}
We concentrate on gravitational studies required in two
main encompassing themes: {\em TeV-gravity} and {\em Holography}. The
former arises as a natural extension of gravity driven, at a very basic
level, by attempts to explain the hierarchy problem.
As first noted in
\cite{ArkaniHamed:1998rs,Antoniadis:1998ig,Randall:1999ee,Randall:1999vf},
this problem can be addressed by considering large volumes
and/or warping within higher
dimensional scenarios~\cite{Giddings:2001yu,Giddings:2007nr}.
Brane world models achieve this goal in a
rather natural way; all fundamental interactions except gravity
are confined to a 4-dimensional brane while gravity
permeates through a bulk - higher dimensional space. Unification of
energy scales can take place at $\simeq$ TeV, a regime which can be
probed at the most powerful currently available accelerators which
may thus open up a window to
new physics~\cite{Giddings:2007nr}. Furthermore,
these TeV-scale gravity scenarios can be realized within String Theory
and thus are tied to a strong prospective quantum theory of gravity. Thus,
we could be on the verge of detecting clues from quantum gravity
behavior in the near future.

Holography, and in particular $AdS$/CFT
dualities~\cite{Maldacena:1997re}, refer to a remarkable relation
between $D$ dimensional Field Theories and gravity in $AdS$ in $D+1$
dimensions~\cite{Maldacena:1997re,Polchinski:2010hw,McGreevy:2009xe}.
It is conjectured that this relation can be exploited to understand
physics on either side of the relation with respect to the other. It
provides ways to understand, or re-interpret,  the behavior of condensed
matter, plasmas, etc. in terms of $AdS$ black holes interacting with
particular fields. For the special case of $D=4$ ``regular physics'', the dual
corresponds to $AdS$ $D=5$ gravity.  Furthermore, it is commonly believed
that the dualities so far uncovered may be
first examples of a much broader class of gauge/gravity dualities.\\

Both themes thus involve understanding gravity in dimensions
higher than $4$, and possibly additional higher curvature corrections
depending on the specific regimes under study. Unfortunately our
understanding of gravity in  $4$ dimensions does not necessarily
help in providing a good intuition of the expected behavior
in higher dimensional settings. Indeed, the non-existence
of stable circular orbits for point particles around black
holes\footnote{The ``centrifugal'' potential barrier is inefficient
beyond $D=4$.}, non-uniqueness of black hole solutions (e.g. see
~\cite{Emparan:2008eg,Emparan:2009vd,Emparan:2010sx,Horowitz:2011cq})
and generic violation of cosmic censorship~\cite{Lehner:2011wc} are just
a few examples illustrating that gravitational phenomena in higher dimensions
are richer than in $D=4$. Furthermore, just as in $D=4$, understanding
generic behavior of solutions to Einstein equations represents a formidable
challenge requiring intense efforts. Coordinated studies are
therefore necessary to gain such intuition and make direct contact with possible
observational efforts. In the following we primarily concentrate on issues
relevant to the numerical construction of $D>4$ spacetimes and highlight
possible research directions. Whenever possible we will make direct
connection with the status of numerical relativity and lessons learned
from work in $D=4$ asymptotically flat spacetimes relevant to these goals.

\subsection{Challenges ahead} \label{challenges}
In the past decade, the field of Numerical Relativity has made great
strides in modeling $D=4$ asymptotically flat spacetimes and is now
able to solve strongly gravitating/highly dynamical scenarios. A
considerable portion of this was driven by the goal of producing
answers to guide detection and analysis strategies for gravitational
wave astronomy (see~\cite{Aylott:2009tn} and references cited therein)
although progress has been achieved on many fronts: gravitational waves,
cosmology, fundamental studies of gravity, astrophysics, etc.
For further details, we refer the reader to some reviews
\cite{Reula:1998ty,Lehner:2001wq,Berger:2002st,Pretorius2007a,ShibataLR,
Sperhake:2011xk,2011PThPS.190..282Y,2011PThPS.189..269Y}
and textbooks~\cite{2008itnr.book.....A,bona2009elements,2010nure.book.....B}
on the subject.

This success has been achieved by the combination of several ingredients:
\begin{enumerate}
\item A good understanding on how to frame the problem in a manner that
is both physically complete and mathematically well posed.
\item The availability of well understood approximations to describe
the state of the solution for stages where the dynamics is relatively
mild. This helps not only in constructing physical initial stages but also
in finding an (approximate) description of  systems where non-linear
effects have little relevance (through e.g. Post-Newtonian or Effective
Field Theory approaches, see also Sec.~\ref{sec:approximation}).
\item The knowledge, or expectation, of the expected late-time behavior
of the system.  Understanding that ultimately a stable Kerr black hole
would most likely describe the asymptotic solution motivated coordinate
and boundary conditions to aid the numerical implementation. Furthermore,
it allows one to devise an effective perturbative description of the
solution (through BH perturbation theory), which in turn alleviates the
computational cost to obtain a solution and improves our ability to understand
the expected generic behavior.
\item A well understood and robust set of numerical techniques to
discretize Einstein equations.
\item Adequate computational resources to deploy the implementation,
and obtain results in a reasonable time for further tests as well as
fine-tuning the implementation and obtain relevant solutions.
\end{enumerate}

The points above are generic requirements for any implementation, and
as we argue in the following discussions, the status of these points in
$D>4$ is less than what would be desired --even with respect to point
(i) above. We next discuss several relevant questions related to the
particular themes we have in mind. Some of these are common to both
broad problems, while others are tied to just one.

\subsubsection*{TeV-scale gravity.}
This putative description of nature does not have a unique theoretical
predictive formulation, rather it is a generic framework where different
theories can be formulated such that they naturally address the hierarchy
problem and possibly connect with TeV physics. While this facilitates the
broad theoretical discussion towards achieving a fundamental description
of nature, at a practical level it makes it difficult to choose which
option to concentrate on. Furthermore, in many cases the full problem can
not be completely defined at a physical level, let alone a mathematically
rigorous one. Broadly speaking one has, in a sense, too many (!) options:
General Relativity, modifications of GR or alternative gravitational
theories, not to mention that, in addition, the treatment of some basic
phenomena might involve important quantum corrections.
Among the set of possible theories, GR is
unambiguously defined, can be shown to define well posed problems and
provides a unique framework to work on.  As far as
alternatives to GR are concerned, it is unclear which theory to pursue
(some are incomplete: e.g. have higher
curvature corrections which have not been explicitly presented, others
are either known to be ill-posed --but are actively pursued
nonetheless-- or
not yet fully understood regarding
the extent to which they might be mathematically
sound (e.g.~\cite{Motohashi:2011ds,Garfinkle:2010zx,Cardoso:2009pk});
see Section \ref{sec:alternative} for a discussion on this subject.
Naturally, 
well posedness is a necessary condition for
a successful numerical implementation.
Several options remain valid,
all with a high degree of complexity and associated computational cost
in practical applications. 
Therefore,
the particular physical situations considered and theory to describe them
must be judiciously chosen to maximize their practical impact. Even a
rather coarse understanding of the possible physics outcome will require
time and concentrated efforts.  An alternative, complementary way  could
be envisioned taking a page out of the ``parameterized post-Newtonian''
approach (PPN)~\cite{Will:2005yc,Will:1989sq} developed within $D=4$
gravitational theories, see Section~\ref{sec:approximation}.
An analog framework for higher dimensional
gravitational studies would be extremely beneficial. Our current
focus is not to investigate this option, but rather to
discuss the role of numerical simulations can play in this problem.
For concreteness, we will concentrate on General Relativity and the main
systems of interests are those that could be realized in accelerators,
(e.g.~\cite{Giddings:2001bu,Dimopoulos:2001hw,Cavaglia:2002si,Kanti:2004nr,Landsberg:2006mm}).
With a concrete theory in mind, the next step is to define relevant
physical scenarios. In particular, for large enough black holes, one can
expect that General Relativity does provide the generic high-energy
behavior as higher curvature corrections become sub-dominant. Of
utmost interest is the understanding of $D>4$ collisions, with non-zero
impact parameter, in order to improve predictions relevant to TeV-scale
scenarios.  Note that these processes are essentially local, and can
therefore be studied with $\Lambda=0$  (asymptotically flat scenarios).
There remain delicate questions, however, which need to be addressed in
order to define concrete target problems; e.~g.~``which is the
dimensionality of the process?'', ``which other non-vacuum interactions
need to be considered?''; ``how to treat interactions restricted
to the 4D brane as well as possibly incorporate
higher curvature corrections?''
From a physical point of view,
one is interested in understanding
high speed collisions or scattering, as well as the behavior
(stability, decay, etc) of black holes possibly resulting from
highly energetic collisions. The latter bears strong relevance to the
former problems, because understanding the stability of possible black holes
cannot only bring important clues about the physical behavior of the
system, but also help in achieving a stable implementation.

\subsubsection*{Holography.     }
Here the situation is  more concrete. The theoretical framework
is most commonly defined with ${AdS}_5 \times S^5$ (though
other scenarios including asymptotically flat spacetimes are under
study~\cite{Guica:2008mu}). The presence of the $S^5$ symmetry helps in
reducing the practical dimensionality of the problem. It is crucial to
understand the limiting cases of the correspondence where the gravitational
side is captured by classical GR in $AdS$ (possibly coupled to gauge
fields) and identify what states on the CFT side they correspond to.
Interestingly, most questions of interest in the CFT side, involve
understanding black hole solutions in $AdS$. Rigorous results about
the stability of generic black holes are lacking with the exception
of Schwarzschild $AdS$ black holes interacting with a scalar field in
spherical symmetry~\cite{Holzegel:2011rk}.
The presence of the $AdS$ time-like boundary allows for fields to
propagate away from the central region, bounce off the boundary and
return to interact. As a result, {\em asymptotically flat intuition}
does not translate to the $\Lambda < 0$ case. Indeed, as recently shown
in~\cite{Bizon:2011gg} and further argued in~\cite{Dias:2011ss}, a
complex ``turbulent-like'' phenomenology arises which renders pure $AdS$
nonlinearly unstable to black hole formation regardless of how weak the initial
perturbation is. This is in stark contrast to the known stability of
Minkowski in asymptotically flat scenarios~\cite{1993gnsm.book.....C} (see
also,~\cite{Lindblad:2004ue}). This illustrates how richer phenomenology
can be expected, and might carry with it interesting ties to diverse
CFT behavior. Indeed, since the proposal of this duality, a plethora
of work has been presented indicating connections across many areas
of physics: fluids, plasmas, condensed matter theory, etc.  (see e.g.
~\cite{McGreevy:2009xe,Gubser:2009fc,Schafer:2009dj,Janik:2010we,Polchinski:2010hw,Horowitz:2010gk,Bernamonti:2011vm,Hartnoll:2011fn}
for some recent discussions). Thus, it is safe to expect no shortage of
interesting connections.

With such a large number of exciting physical arenas involved, it would be
useful to understand where numerical studies can make the highest impact.
In order to address this question, it is important to comprehend which ``real world''
applications (e.g. quark-gluon plasmas
\cite{CasalderreySolana:2011us,Mateos:2007ay},
superconductors \cite{Hartnoll:2009sz,Hartnoll:2008vx})
or generic physical behavior (e.g. turbulence \cite{Bizon:2011gg}) 
can be accurately captured through a CFT model so as to study essential physical features of the system one
is trying to model and identify those that can be studied
within a reduced dimensionality (see section \ref{dynamical}).

A particularly delicate issue for a numerical implementation concerns the
$AdS$ boundary. Indeed, as opposed to the asymptotically flat case
where numerical implementations typically deal with boundaries with
a combination of techniques to ensure they do not play a role in the
dynamics, here the situation is markedly different.  Not only would the
boundary be in causal contact with the interior domain, but also the
field behavior in its vicinity is of special interest.  Indeed, one
goal required from a numerical implementation is to allow for extracting
the asymptotic behavior and connecting with the boundary theory.  As we
discuss later the mathematical understanding of how to pose correct
boundary conditions is lacking. Further details can also be found in Section~\ref{sec:holography}.

\subsection{Understanding strongly gravitating, dynamical systems in $D>4$}
\label{dynamical}
The problems of interest concern energetic events involving
black hole formation, black hole interactions and/or highly perturbed
black holes. For specific limits, e.g. ultra-high-speed collisions,
particular approximations have proved quite useful in exploring the
phenomenology of the system~\cite{Eardley:2002re}. Here fields of
particles can be described by the Aichelburg-Sexl solution and show
black holes form for a range of impact parameters.
For more general scenarios, perturbative studies might be
unable to capture the sought after behavior correctly, especially during
highly nonlinear stages. Solutions are thus to be sought within the full
theory which requires numerical relativity.

Obviously, the study of general scenarios in higher dimensions
are computationally more demanding. At a rough level,
the computational cost scales  as $C^{D}$ for general situations in
$D$ dimensions. Putting the requirements into perspective, fairly generic $D=4$
scenarios are currently routinely studied and typical vacuum (binary
black hole) simulations demand a few weeks of continued run on several
dozen processors\footnote{Details depending on the accuracy
requirements; non-vacuum scenarios can take significantly
more depending on the complications introduced by additional physical
ingredients involved.}. Extrapolating from this observation, one can
argue that general scenarios could be studied in $D=5$ with current
and near future resources.  However, this will come at a significant cost,
especially when progress in this front might require, as has been
the case in $D=4$, some experimentation to achieve a robust code. Thus, higher
dimensional cases in generic settings, will in practice be out of the
question for a long time. Therefore, it is safe to expect that Numerical
Relativity will have its highest impact in studying scenarios where
certain symmetries can be adopted, thus reducing its dimensionality to
more tractable 2 and 3 spatial dimensions. Particularly relevant cases
are those in which $SO(D-2)$ and/or $(D-2)$ planar symmetries can be assumed
such that the dynamics of interest takes place in a dimensionally reduced
setting~\cite{Emparan:2008eg}.

To date, a small number of numerical relativity efforts have been carried out
beyond spherically symmetric scenarios to study specific
questions in higher dimensional spacetimes.  Within large extra
dimension scenarios, these have concentrated on examining black hole
instabilities~\cite{Shibata:2009ad,Lehner:2010pn,Shibata:2010wz} and high
speed collisions~\cite{Zilhao:2011zz,Witek:2011zz,Witek:2010xi,Witek:2010az,Okawa:2011fv}. These implementations
have mostly followed the two successful approaches within the Cauchy
formulation of Einstein equations used in $D=4$: the generalized
harmonic and BSSN formulations together with gauge conditions that are
direct extensions of those employed there. They have further
taken advantage of generic computational infrastructure developed to handle
parallelization and adaptive regridding so as to ensure an efficient
usage of computational resources (e.g. like the publicly available
Cactus/Carpet, PAMR, HAD ~\cite{Cactus,PAMR,HAD}).  Within holographic studies,
published works of black hole formation have relied on the characteristic
formulation of  Einstein equations~\cite{Chesler:2009cy,Chesler:2010bi}
and ongoing work is also exploiting the generalized-harmonic
approach \cite{Bantilan:2012vu}.

Interestingly, within the Cauchy approach, directly exploiting
symmetries (say in $p$ of the dimensions) assumed at the onset, and
defining a reduced problem has given difficulties. These can be traced
to the fact that Cartesian coordinates satisfy the harmonic condition
while spherical ones do not, giving rise to non-trivial sources which
couple numerical errors to a lower order. There is no fundamental
impediment to deal with this issue, and successful runs have been performed even for five dimensional spacetime \cite{Zilhao:2011zz,Witek:2011zz,Witek:2010xi,Witek:2010az}.
On the other hand, implementing symmetries in an effective way following the ``cartoon
method'' (where the problem is treated formally as $D$
dimensional, but the equations are integrated in a $(D-p)$ sector and
derivatives off this sector are accounted for via the symmetries) have
provided robust implementations in a rather direct, though perhaps not
elegant way~\cite{Shibata:2009ad,Lehner:2010pn,Shibata:2010wz}. This
observation stresses that more work at the foundational level  is still
required to translate the success obtained in $D=4$ to higher dimensional
settings. Beyond these issues, we next discuss some particular
points of relevance to our two encompassing themes.

{\em In the context of TeV-scale gravity scenarios}, translating the
success of $D=4$ simulations is in principle relatively direct;
in vacuum scenarios, in particular, the intrinsic computational problem is
quite similar and existing computational infrastructure can be readily
exploited.  In essence, one deals with an {\em initial value problem}
and computational boundaries need just to be dealt with in a stable manner
and placed sufficiently far from the region of interest so they do not
causally influence it. After dealing with the dimensionally reduced
problem with either option above, one can fine tune the implementation
(adjusting gauge conditions, refining algorithms, etc) and begin
studying problems of interest provided the correct physical ingredients are incorporated in the model.\\

{\em In the context of $AdS$-CFT dualities}, the above picture
is more complex as the problem is intrinsically an {\em initial boundary
value problem}. The $AdS$ time-like boundary is causally connected with
the bulk region and must be carefully treated as it plays a central
role in the solution sought.  Understanding how to deal with time-like
boundaries in $D=4$ in the absence of a cosmological constant required
major theoretical efforts which culminated in a series of mathematically
sound options guaranteeing the well-posedness of the underlying
problem~\cite{Kreiss:2007zz,Reula:2010yt}. An analog of such work is
absent in $AdS$ (even in $D=4$) which is rendered more delicate due to the
diverging behavior of the spacetime metric at the boundary.
We note, however, the derivation of boundary conditions for the linearized
field equations in $AdS$ \cite{Ishibashi:2004wx} and studies of the
fall-off behaviour of the metric in particular coordinate systems
\cite{fefferman1985,Henneaux:1985tv}.
Extension of this work towards establishing well-posedness of the
full equations will be of vital importance
for a robust numerical treatment of
general problems. In the meantime however, interesting advances have been
presented exploring the characteristic formulation of General Relativity
--where the spacetime is foliated by incoming null hypersurfaces
emanating from the $AdS$ boundary--~\cite{Chesler:2009cy,Chesler:2010bi}
(which is particularly robust,
a property that has been
observed in $D=4$ as well) but more restricted in applicability as caustics will render the
coordinate system employed singular. At present
there is no general purpose infrastructure to ensure efficient
parallelization and dynamic regridding for characteristic approaches,
though a proof-of principle work has demonstrated there is no fundamental
obstacle for developing one~\cite{Pretorius:2003wc}.  Within a Cauchy
approach --where the spacetime is foliated by spacelike hypersurfaces--,
strategies have been specifically tailored for special cases, see Section~\ref{sec:holography}.

Regardless of the nature of the foliation adopted, a related question
concerns the boundary topology.  It is well known that within the global
5D-$AdS$ spacetime (with boundary topology $R\times S^3$) a Poincar{\'e} patch can be
defined through a suitable transformation.
The boundary of this patch is $R^4$.  From the perspective of aiming for a numerical implementation,
these two options might bring up non-trivial issues related to both
the possibility of achieving a robust implementation and its application
to generic problems. Namely, if a global $AdS$ picture is adopted, can all
physics be extracted in standard fashion through a Poincar{\'e} patch after a suitable coordinate transformation?
As a matter of principle, a calculation performed in global $AdS$ can
always capture the physics of a calculation in the Poincar{\'e} patch,
although in practice it may be a rather inefficient and involved manner
of doing so. The converse, however, could not always be possible. In Euclidean
space, the global and Poincar{\'e} patches are, as mentioned above,
convertible into each other through a rather straightforward coordinate
transformation in a
unique way. However, in Lorentzian spacetimes the situation is more
subtle (one can presumably ascribe this to the different possible
prescriptions for how the continuation from Euclidean to Lorentzian is
performed). Excitations or objects evolving in global $AdS$ may cross the
Cauchy horizon of the
Poincar{\'e} patch, entering or leaving it (their
energies, measured in Poincar{\'e}-patch time, are
redshifted to zero as they get to the Cauchy horizon).
As long as the relevant physics is occurring entirely inside
the Poincar{\'e} patch, this is not problematic, but
for phenomena where properties on global $AdS$ are required,
Poincar{\'e}-patch evolutions appear to be
ill-suited.
Although in
principle analyticity should allow for recovering Lorentzian-time global-$AdS$
correlation functions from Poincar{\'e}-patch ones, one must have a detailed
understanding of the prescriptions involved and the approach 
is likely unsuitable for numerical results.
Moreover, relating the Poincare picture to that of the global picture
may require infinite resolution, and thus be difficult in numerical
approaches.

Alternatively, are particular conditions required in the boundary
treatment of global $AdS$ to ensure this is the case? Furthermore, some
applications require ``operator insertions'' in the CFT that correspond
to particularly ``deformed'' $AdS$ boundaries. Here again, all issues  with
respect to fall-off behavior, boundary regularization, well-posedness of
the resulting problem, need also to be investigated for these spacetimes. It
is important to stress here that some of these questions  (such as
Poincar{\'e} patch or not) may have simple answers from the physics
perspective, but they might introduce delicate numerical issues that
may also guide the approach to solutions.

\subsubsection*{Fields, what fields?}
An issue related to both broad themes concerns the choice of fields to be
included for studying relevant scenarios. For instance:
In the context of $AdS$/CFT, the inclusion of a gauge field is natural and
well-motivated, as it allows the study of the effect of non-zero chemical
potential, i.e., finite density, in the boundary theory. Scalar fields
with suitable potentials are also of interest and have been studied
to model in a phenomenological manner the effects of confinement
(``soft-wall'' models).

\begin{itemize}
\item For TeV-scale gravity, considering charged collisions, or those
involving other fields, will allow an exploration of which other observational
signatures could be sought after in accelerator experiments. In $D=4$,
simulations involving scalar and vector interactions are well under
control (e.g.~\cite{Choptuik:2009ww,Palenzuela:2006wp,Neilsen:2010ax}).
In higher dimensions, however, these fields would be confined
to branes and the treatment becomes considerably more complicated.
\item In the context of holography,
studying the interaction of black holes with other fields 
is important for exploring dualities within a broad physics
scope. This includes the ``condensation'' of fields outside black holes, which has proven 
important for studies of superconductors and generic (holographic) condensed matter systems.
Of particular importance, for instance, is to understand the
global behavior driven by super-radiance of scalar or even tensor fields \cite{Cardoso:2004hs,Kunduri:2006qa} and its final outcome, likely,
a hairy black hole or non-trivial black hole solutions.
\end{itemize}

While exciting work is proceeding to address problems of interest, there is undoubtedly still plenty of room for resolving fundamental
questions that will help in the construction of robust and general methods to
simulate relevant systems and to extract useful physics from them.

\subsection{Targets of opportunity}
As mentioned, gravity in higher dimensions brings about several new
delicate issues, some of which can be addressed via numerical simulations
while others must be taken into account analytically
in order to achieve a
robust implementation.  Indeed, essentially every single point discussed
in Sec.~\ref{challenges} would benefit from further work: while some
intuition from $D=4$ can be brought to higher dimensions for guidance,
the richer phenomenology of gravity in $D>4$ indicates that such intuition
can only help within limits.
A few simple examples of radically different behavior are
as follows: non-existence of stable circular orbits even when ignoring
back-reaction; non-uniqueness of black hole solutions (distinct solutions
are known with the same asymptotic charges); the possibility of naked
singularities arising in generic scenarios (already shown for the GL
instability); in brane-world scenarios gravity sees the bulk while
other interactions are restricted to the brane. In addition, there
are important quantum issues not covered here, such as the behavior of
Hawking evaporation and non-perturbative quantum gravity effects.  On the
one hand, these issues make the understanding of these
systems more difficult; on the other hand, they represent targets of opportunity. Indeed,
exploring them will certainly provide interesting new challenges in their
own right.  While the landscape of possible systems is still largely
unexplored, from a practical standpoint, problems can be divided as:
\begin{itemize}
\item Mathematical. Understand the correct setting for guaranteeing a
well posed initial boundary value problem in $AdS$.
Understand black hole topologies that might arise in higher
dimensional contexts~\cite{Galloway:2011zp}.
\item Mathematical/Physical.
Very basic issues about the stability of higher-dimensional black holes
remain unresolved and at present we can only expect to  address them
through numerical studies. Results
thus obtained can
guide us towards a rigorous
path to establish general results from a mathematical point of view.
As discussed earlier, this is particularly important for the scenarios
in which black holes can be formed at colliders.  Two outstanding
open (and related) questions in the context of vacuum black holes (with
$\Lambda=0$) are: (i) black rings with large angular momenta are unstable,
but is there a window of stability for (5D) black rings at moderate
values of the spin? (ii) for given values of $M$ and $J$, is there more
than one stable black hole?  If the answer is yes, it would have
consequences, for instance, for black hole production at colliders. If not,
this would allow us to recover a notion of uniqueness: the only neutral,
asymptotically flat black holes with a connected horizon, would be
Myers-Perry black holes with angular momentum below a certain bound.

More generally, numerical analysis, in a less intensive manner (``soft
numerics'') is necessary for obtaining information about the space of black
hole solutions in higher-dimensions that cannot be obtained through
other methods. A good example of this is the progress in the past decade
in understanding black holes and black strings in Kaluza-Klein theory. Very
similar problems arise in any other higher-dimensional theory of gravity,
either in asymptotically flat vacuum, or with a cosmological constant or
additional fields.

\item Physical. List scenarios relevant for TeV-gravity and Holography
in both vacuum and non-vacuum spacetimes.
To this end, describe desired initial conditions
required to study the future development of the solution. Furthermore,
investigate which particularly interesting cases are amenable to a
reduced dimensionality description.  Develop methods that could shed
further light on the linearized stability of many black hole solutions. For
instance: recognize mappings between different solutions and known cases
(as in the case of mapping close horizon geometries to the GL instability
e.g.~\cite{Emparan:2003sy,Dias:2010eu,Dias:2009iu}); extend, and comprehend possible limitations
of thermodynamical arguments (e.g.~\cite{Gubser:2000ec,Figueras:2011he,Dias:2010eu});
construct suitable approximating schemes able to capture the
system's behavior in appropriate limits (e.g. extensions of PN and
EFT~\cite{Kol:2007bc}, blackfolds~\cite{Emparan:2009at}, etc.)
\item Numerical. Several open problems should be solved numerically,
both to obtain
static/stationary solutions or particular initial data as well as studying
the dynamical behavior of relevant cases. In the following sections we
discuss some particularly interesting issues on this front.
\end{itemize}

\subsubsection*{Initial data: stationary/static solutions and perturbations
thereof.}
Of particular interest are black hole solutions; as mentioned,
these can form as a result of high-speed collisions or are
intrinsically important within $AdS$-CFT dualities. A number of
solutions are known, but generically their stability behavior is not
understood (see further discussions in section \ref{BHs}). In
some cases, important headway has been made by entropic arguments
and recognized mappings to known instabilities (most notably the Gregory-Laflamme (GL)
instability~\cite{Gregory:1993vy,Gregory:1994bj}).  For those, recent
work uncovering what the non-linear dynamics of the system might be, gives
an indication of what to expect in general. 
\begin{it}
With the exception of the GL instability,
present options (short of performing a full non-linear study) are
linearized perturbation analysis or considering scenarios describing black
hole perturbations violating Penrose's inequality~\cite{Figueras:2011he}.
\end{it}
To date, linearized stability of many of the known black hole solutions
is not yet known.  The question is further complicated as many
solutions for interesting cases are only known numerically. Indeed,
many situations of interest relate to static or stationary black hole
solutions~\cite{Wiseman:2011by} whose derivation requires significant
work at analytical and numerical levels.

\subsubsection*{Dynamical behavior.}
Understanding of the full non-linear behavior of black-hole
perturbations at both classical and quantum levels is paramount
for establishing a connection with possible observations in the framework of TeV-scale gravity models as
well as elucidating holographic connections with equilibration in the
dual picture. Furthermore, studies of dynamical scenarios in higher
dimensional gravity will help in building up intuition and guide
further work. For instance, numerical simulations have already shown
(see also~\cite{Yoshino:2011zz}):
\begin{itemize}
\item That negative energy ``bubbles of nothing'' do not give
rise to naked singularities, and initially expanding ones continue
doing so. Furthermore, that gauge fields can significantly affect the
dynamics and give rise to a static bubble solution dual to black
strings~\cite{Sarbach:2003dz,Sarbach:2004rm}.
\item That large static black holes exist in $AdS_5$ with the Schwarzschild
black hole as the boundary metric~\cite{Figueras:2011va,Figueras:2011gd}
thus disproving a conjecture that no such solutions are
admitted~\cite{Tanaka:2002rb,Emparan:2002px}.
\item That a higher dimensional class of unstable black holes --black
strings--\cite{Gregory:1993vy,Gregory:1994bj} display rich dynamics
leading to a self-similar behavior that ultimately gives rise to naked
singularities~\cite{Lehner:2010pn,Lehner:2011wc} in rather generic
conditions. Thus cosmic censorship does not hold beyond $D=4$.
\item The unstable behavior of rapidly spinning black
holes~\cite{Emparan:2003sy}, which can lose enough angular
momentum through gravitational waves so as to cross to the stable
branch~\cite{Shibata:2009ad,Shibata:2010wz}.
\item That the amount of energy radiated in $D=5$ head-on black hole
collisions agrees well with the value obtained from (extrapolations of)
linearized, point-particle calculations~\cite{Witek:2010az,Berti:2010gx}.
\item That within the holographic picture, the collision of gravitational
shock waves in $AdS$ gives rise to a dynamical behavior consistent with
that expected from hydrodynamics in the boundary theory. Furthermore, the
behavior in the gravity sector was exploited to compute the time required
for thermalization in the system \cite{Chesler:2010bi}.
\end{itemize}

These are just a few examples of interesting physics that can be extracted
from the dynamical behavior of relevant systems. A priori there is a
vast number of interesting problems and this number will likely grow
as new, and probably unexpected, behavior is uncovered in the
process. Obviously prospects for exciting physics lie ahead. Indeed,
within large extra dimensions, a thorough understanding of how a newly
formed --but unstable-- black hole decays to its eventual fate, what
that fate is and how it depends on dimensionality can have profound
observational implications and/or signal difficulties in prospective
models. Furthermore, the influence of the strongly gravitating/highly
dynamical regime explored by such a process on additional fields
could induce tractable signals that can provide important additional
observational channels.

Within holographic studies for static situations there is a clear
prescription on the relationship between the CFT thermal state and
properties of the BH horizon. However, for dynamical scenarios --related
to both formation and stabilization of black holes-- is there a sensible,
unique way to define such a map in a dynamical situation? Furthermore,
what is the preferred way to connect CFT and horizon properties as a
function of time? The answer to this question necessarily ties to the
choice of a preferred slicing, which need not be the most convenient
one for the numerical implementation.  Additionally, during the
dynamical stage complex dynamics might be driven by the super-radiance
``instability''. However, to date, a study of such scenario even in $D=4$
is absent. Related work has presented possible end-state solutions
describing hairy black holes~\cite{Dias:2011at,Stotyn:2011ns,Basu:2010uz,Gentle:2011kv,Dias:2011tj},
though their stability is not yet established nor whether they are
indeed attractor solutions. However, the super-radiance instability
need not be the only or main effect driving the dynamics. For
instance, recent work indicating `turbulent-like' behavior in $AdS$
gravity~\cite{Bizon:2011gg,Dias:2011ss} highlights surprises might still
be lurking in dynamical scenarios beyond what might be understood at
low perturbation orders. Understanding these phenomena and their possible
dual on the field theory side can shed light on key behavior or raise
further questions. For instance, this turbulent-like behavior takes
place on the gravity side in all dimensions.  However, the Navier-Stokes
equations (appearing in to lowest order on the field theory side),
display a behavior markedly different in $D=3$ with respect to higher
dimensions. This reveals that a different phenomena is awaiting our
understanding in the context of strongly coupled QFT.

\subsubsection*{Drawing a path from analogies to 4D.}
As mentioned above, many cases of interest involve understanding
the interaction of black holes in a fully non-linear framework. It is
important to recall once more that simulations provide information about
one single, isolated black-hole system per run. Having a sufficient
number of such simulations available, certain phenomenological models
could be constructed to capture relatively generic behavior with respect
to certain regions of the parameter space. Undoubtedly this would be a
time-consuming task. An alternative (or complementary) approach has been
pursued successfully in $D=4$. Namely to employ different approximations
to understand the system in separate (early and late) regimes and
exploit numerical simulations to bridge the gap in between.  Techniques
used for this purpose are Post Newtonian (and related) approximations
when the compact objects move at slow velocities and
perturbation theory around a suitable
black hole solution for the post-merger (or black-hole formation) stage
(e.g~\cite{Pullin:1999rg,Campanelli:2005ia,LeTiec:2009yf,review:2007mn,Pan:2009wj,Santamaria:2010yb}
for $D=4$ studies
 and~\cite{Yoshino:2005ps} for a first step in this direction for
 $D>4$.). Consequently, a judicious phenomenological
match between the two phases, motivated and further tuned via gradual
generation of numerical solutions, can provide for an efficient way to
encode the system's behavior.  This approach should be given consideration
in higher dimensional scenarios.  However, to this end the following
issues must be kept in mind:
\begin{itemize}
\item PN and EFT approximations must be developed to the
appropriate order. In particular to incorporate radiation-reaction
effects. Currently this is only available to first order (and not in
the $AdS$ case)~\cite{Kol:2007bc}.  It is important to note that the
order at which this can be done without internal effects playing a role,
depends on dimensionality. Nevertheless, where possible, knowledge of
the trajectories can be directly exploited in obtaining reasonable
approximations to the spacetime metric by suitable superpositions (see
e.g.~\cite{Bishop:1997my,Marronetti:2000rw,Sarbach:2001tj,JohnsonMcDaniel:2009dq}).
\item Perturbations of black holes only make sense if the stability of
the black hole is understood. As mentioned in many cases this is still
to be addressed.
\end{itemize}

For a detailed discussion of the use of approximation methods for
$D \ge 4$-dimensional black-hole spacetimes see Sec.~\ref{sec:approximation}.

%% file: BHs.tex
\section{Higher-dimensional black holes} \label{BHs}
\begin{center}
Coordinator: Harvey Reall
\end{center}

\subsection{Motivation}
This
section will discuss classical properties of stationary black hole
solutions of the vacuum Einstein equation in higher dimensions.

The review article \cite{Emparan:2008eg} listed several motivations for the study of black hole solutions in more than four spacetime dimensions:
\begin{enumerate}
\item
Statistical calculations of black hole entropy using string theory. This was first achieved for certain 5-dimensional black holes and later extended to 4D black holes. Each entropy calculation is a check on the theory, irrespective of the dimension. Hence the study of higher-dimensional black holes is a worthwhile contribution to developing a theory of quantum gravity.
\item
The gauge/gravity correspondence relates the properties of black holes in $D$ dimensions to strongly coupled, finite temperature, quantum field theory in $D-1$ dimensions. This provides a way of calculating certain field theory quantities which cannot be determined by any other method.
\item
Certain ``large extra dimensions'' scenarios predict that microscopic higher-dimensional black holes might be formed at the LHC. However, LHC results discussed at the workshop give no evidence in favour of ``large extra dimensions'' scenarios. It was also emphasized that if black holes were formed at the LHC when run at higher energy, they would be not be in a semi-classical regime, so quantum gravity would be required to study them. Therefore it seems that this motivation for the study of classical higher dimensional gravity is lessened. 

\item
Higher-dimensional black hole spacetimes might have interesting mathematical properties. For example, analytically continued versions of black hole solutions have been used to obtain explicit metrics on compact Sasaki-Einstein spaces.
\item
Just as it is valuable to consider quantum field theory with field content different from that of the Standard Model (or any conceivable extension), it might be worthwhile considering higher dimensions when studying black holes. For example, there might exist explicit higher-dimensional solutions that provide a clean example of some effect in GR. A nice example of this is the frame-dragging effect exhibited by the ``black Saturn'' solution \cite{Elvang:2007rd}. Perhaps there are examples in which an exact calculations in higher dimensions can be used to check a calculation that has to be done perturbatively in 4D.
\item
Some things are simpler in higher dimensions. For example, in 4D the asymptotic symmetry group of null infinity in an asymptotically flat spacetime is the infinite-dimensional Bondi-Metzner-Sachs group. However, in higher dimensions it is simply the Poincar\'e group \cite{Hollands:2003ie,Tanabe:2011es}. This makes it possible to define angular momentum at null infinity in higher dimensions \cite{Tanabe:2010rm}. In 4D this appears to be difficult.

\end{enumerate}

The focus of this article will be on stationary black hole solutions and their properties. Time-dependent processes are of great interest for the applications but will only be touched on here (see Sections \ref{sec:nr} and \ref{sec:holography} for further details on this topic).

\subsection{State of the art}

\subsubsection*{Explicit solutions.}

We restrict this discussion
to the vacuum Einstein equations with vanishing cosmological
constant. There are two families of explicit asymptotically
flat black hole solutions with a connected horizon. There are
the Myers-Perry solutions \cite{Myers:1986un}, known for any
spacetime dimension $D$, and the 3-parameter black ring solution
\cite{Emparan:2001wn,Pomeransky:2006bd}, for $D=5$. The MP solutions
have horizons with topologically spherical cross-section. The black
ring solution has topology $S^1 \times S^2$.

There are also explicit solutions with disconnected horizons. The ``black Saturn'' solution \cite{Elvang:2007rd} describes a black ring with a MP black hole sitting at the centre of the ring. There are solutions involving a pair of black rings, e.g., the ``black di-ring'' \cite{Iguchi:2007is}. 

The MP solutions have been generalized to include a cosmological constant for $D=5$ \cite{Hawking:1998kw} and $D \ge 6$ \cite{Gibbons:2004uw}. These solutions describe rotating, topologically spherical black holes in an asymptotically (anti-)de Sitter spacetime. 

\subsubsection*{Classification.}

It has been shown that a static, asymptotically flat, black hole must
be described by the Schwarzschild solution in any number of dimensions
\cite{Gibbons:2002av}.

Hawking's topology theorem has been generalized to higher dimensions \cite{Galloway:2005mf}, with the conclusion that a cross-section of the event horizon of a stationary black hole must be a positive Yamabe space, i.e., it must admit a metric with positive Ricci scalar. In 4D this implies spherical topology. In 5D, it restricts the topology to a connected sum in which each component is either $S^1 \times S^2$ or a quotient of $S^3$. For $D>5$ there are many more possibilities. 

Hawking's rigidity theorem asserts that a stationary black hole
must admit a rotational symmetry. This has been extended to higher
dimensions \cite{Hollands:2006rj}. But it guarantees only one rotational
symmetry whereas all known explicit solutions have multiple rotational
symmetries. Combining the rigidity and topology theorems restricts
further the possible topologies in 5D \cite{Hollands:2010qy}.

For $D=5$, one can classify stationary black holes with two commuting
rotational symmetries according to their ``rod structure'': there exists at
most one black hole with given mass, angular momenta, and rod structure
\cite{Hollands:2007aj}. If one also assumes $S^3$ topology then the black
hole must be a Myers-Perry solution \cite{Morisawa:2004tc}. Solutions
with lens space topology are consistent with this classification. It
is not known whether such solutions exist (they do not appear in the
blackfold approach discussed below).

Related to the problem of classification is the characterization of the
phase space of black hole solutions: what are the different families of
black hole solutions that exist, and how they branch-off or merge at
different points in the phase space. Even if we do not have a complete
classification of all possible black holes, one would like to know how
the known phases (explicit or approximate) relate to each other, at
least qualitatively.

In this direction, one expects that the main features of the phase space
of neutral, asymptotically flat, higher-dimensional
black holes are controlled by solutions in three different regions:
\begin{enumerate}
\item[(i)] Large angular momenta.
\item[(ii)] Bifurcations in phase space.
\item[(iii)] Topology-changing transition regions.
\end{enumerate}
The regime (i) is captured by the blackfold effective theory
\cite{Emparan:2009cs,Emparan:2009at} discussed below. Regions (ii) are controlled by
zero-mode perturbations of black holes that give rise to bifurcations
into new families of solutions. The initial conjectures about these
points \cite{Emparan:2003sy,Emparan:2007wm} have been confirmed and
extended in \cite{Dias:2009iu}. For the regions (iii) in $D\geq 6$,
ref.~\cite{Emparan:2011ve} has provided local models for the
critical geometries that effect the topology change. The critical
behavior in the
five-dimensional case is qualitatively different.

\subsubsection*{Stability.}

Singly spinning MP black holes in $D \ge 6$ have no upper bound on
the angular momentum $J$ for a given mass. Ref. \cite{Emparan:2003sy}
conjectured that such black holes should be classically unstable for large
enough $J$. Strong evidence for this was found in Ref. \cite{Dias:2009iu,Dias:2010maa},
where it was shown that a regular stationary perturbation appears at
a critical value of $J$. This is believed to be a ``threshold'' mode
indicating the presence of exponentially growing perturbations for
larger $J$.

Ref. \cite{Dias:2010eu} considered the most symmetrical case of MP black holes with odd $D$ and all angular momenta equal. Such solutions have an upper bound on $J$ at which the black hole becomes extreme. It was shown that, close to extremality, there exist linearized perturbations that grow exponentially with time. Ref. \cite{Dias:2011jg} showed how the threshold mode of this instability connects to the threshold mode in the singly spinning case by considering unequal angular momenta.

The studies just described considered only instabilities that preserve the rotational symmetry of the black hole that arises from the rigidity theorem. However, in the singly spinning case, it is known that a non-rotationally symmetric instability appears at a lower value of $J$ than the rotationally symmetric instability. Ref. \cite{Shibata:2009ad} performed a full nonlinear evolution of the Einstein equation starting from initial data describing a singly spinning MP black hole with a non-rotationally symmetric perturbation. For large enough $J$, it was found that the black hole became very asymmetrical, resulting in significant gravitational wave emission, and then settled down to a (presumably stable) MP black hole with $J$ always smaller than some critical value. It was found that this kind of instability is present for $D = 5$ (despite the upper bound on $J$) as well as for $D \ge 6$. 

With a negative cosmological constant, MP-$AdS$ black holes suffer a superradiant instability when $\Omega \ell > 1$ where $\Omega$ is the angular velocity of the horizon and $\ell$ the $AdS$ radius \cite{Kunduri:2006qa}. This instability breaks rotational symmetry. A rotationally symmetric instability can also appear at even larger angular velocity \cite{Dias:2010gk}. 

Refs \cite{Arcioni:2004ww,Elvang:2006dd} presented heuristic arguments indicating a classical instability of ``fat'' black rings. This was confirmed by Ref. \cite{Figueras:2011he}  using an argument based on Penrose inequalities. This instability preserves rotational symmetry. Heuristic arguments also suggest that sufficiently thin black rings will suffer a GL instability, which would break rotational symmetry \cite{Emparan:2001wn}. 

\subsubsection*{Approximate techniques.}

Heuristic arguments suggest that black ring solutions also should exist for $D>5$. These are made more precise by the ``blackfold'' approach \cite{Emparan:2007wm,Emparan:2009cs,Emparan:2009at} which constructs solutions perturbatively in a derivative expansion, valid when the geometry of the solution has a large hierarchy of scales, e.g., a black ring with topology $S^1 \times S^{D-3}$ where the radius of the $S^1$ is large compared to that of the $S^{D-3}$. This approach indicates the existence of black hole solutions with a variety of different topologies, for example a product $S^{p_1} \times \ldots \times S^{p_k} \times s^{D-p-2}$ where $p_i$ are odd with $\sum p_i = p$ and $S^{p_i}$ denotes a sphere of ``large'' radius and $s^{D-p-2}$ a sphere of ``small'' radius. It also indicates the existence of ``helical'' black ring solutions with just the single rotational symmetry predicted by the rigidity theorem, i.e., less symmetry than any known higher-dimensional black hole \cite{Emparan:2009vd}.

Another approximate technique which also points to the existence of new
solutions is to consider perturbations of a known solution. If one finds a
perturbation which is stationary, regular, and does not correspond simply
to a variation of parameters of the known solution then it may correspond
to the bifurcation of a new family of stationary black hole solutions. The
studies of MP perturbations just discussed do indeed provide evidence for
such bifurcations. For singly spinning MP black holes with $D \ge 6$,
a stationary perturbation appears at the critical value of the angular
momentum beyond which the black hole is unstable \cite{Dias:2009iu}. This
is taken to indicate the existence of a new family of ``pinched'' MP black
holes. Further bifurcations appear at larger angular momentum. All of
these perturbations preserve the symmetries of the original MP solution.

The same technique applied to cohomogeneity-1 MP black holes also suggests the bifurcation of a new family of solutions at the critical value of angular momentum beyond which the solution becomes unstable \cite{Dias:2010eu}. However, in this case, the perturbation generically breaks all of the rotational symmetries of the MP solution except for the symmetry guaranteed by the rigidity theorem. In contrast with helical black rings, the new family of solutions would be topologically spherical. Furthermore, counting the number of free parameters in the perturbation suggests that the new solutions should have many parameters (e.g. 70 in $D=9$ whereas the 9D MP solution has only 5 parameters).

In summary: approximate techniques indicate the existence of many stationary vacuum black hole solutions in $D>5$ dimensions. These include solutions with only a single rotational symmetry, solutions with topology different from any known solution, and solutions with many more parameters than the known solutions.  

In asymptotically $AdS$ spacetimes, it appears that things can be even more
complicated. Ref. \cite{Kunduri:2006qa}  suggested the possible existence
of stationary, non-static, vacuum black holes without any rotational
symmetry. Evidence in favour of this was found in Ref. \cite{Dias:2011ss}.

\subsubsection*{Numerical solutions.}

The study of asymptotically flat black holes gives one a stepping stone to understand more complicated theories of gravity relevant for phenomenology or holography. In the asymptotically flat setting the static solutions are simply Schwarzschild and are known to be unique \cite{Gibbons:2002bh}. However in more exotic settings, such as compact extra dimensions or braneworlds, even the static black holes often have a complicated structure.
For example in the simplest toy model, Kaluza-Klein theory, where we are interested in vacuum geometries that asymptote to $\mathbb{R}^{1,3} \times S^1$, there are 3 distinct classes of static solution. As reviewed in \cite{Horowitz:2011cq} these are the homogeneous black string, the inhomogeneous black string and the localised black hole. Only the first is known analytically. The second and third class may be constructed perturbatively in various limits, but generally are only known from numerical computations (the most recent being \cite{Headrick:2009pv}). The inhomogeneous black strings were originally predicted by Gregory and Laflamme \cite{Gregory:1987nb} and are generated from the homogeneous strings by the marginally stable static Gregory-Laflamme perturbation \cite{Gubser:2001ac}. Moving along this solution branch the horizon appears to degenerate to pinch off and change the topology to that of a spherical horizon \cite{Wiseman:2002zc}. 
Kol has argued \cite{Kol:2002xz} that on general grounds a singular cone geometry provides a local model for the part of the horizon where the pinch off occurs, and that one may resolve the cone to pass through to the spherical topology horizons of the localised static solution branch. This localised branch is thought to interpolate between this topology changing point and very low mass spherical black holes that near their horizon approximate 5-dimensional Schwarzschild and were first studied by Myers \cite{Myers:1986rx} and may be constructed in perturbation theory \cite{Harmark:2003yz,Gorbonos:2004uc}. This picture is indeed supported by the numerical evidence \cite{Sorkin:2006wp,Headrick:2009pv}. This topology change of the horizon through a conical transition is also thought to be relevant in understanding the space of asymptotically flat stationary solutions \cite{Emparan:2007wm,Emparan:2011ve}. Further work to improve these numerical solutions and confirm the cone topology change picture is an interesting future direction.

\subsection{Future directions}
\subsubsection*{Explicit solutions.}

The approximate methods discussed above point to the existence of many stationary vacuum black hole solutions in higher dimensions in addition to the known Myers-Perry and 5D black ring solutions. It is of obvious interest to determine such solutions explicitly. This is a long-term goal. To do this, new techniques for solving the Einstein equation in higher dimensions must be developed.

So far, the technique which has led to the most interesting results is the Belinskii-Zakharov inverse scattering method for solving the Einstein equation when there are $D-3$ commuting rotational symmetries, as well as time-translation symmetry. This method led to the discovery of doubly spinning black rings and black Saturn and its generalizations. But these symmetry assumptions are consistent with asymptotic flatness only for $D=4,5$. Furthermore, this method does not work with a cosmological constant. However, if 5D ``black lenses'' exist then this method could be used to find them.

An alternative approach is based on algebraic classification of the Weyl tensor. The Kerr solution was discovered in a search for solutions with a Weyl tensor that, in the Petrov classification, is algebraically special. There is a large literature on exploiting the algebraically special property as a tool for solving the Einstein equation in 4D. The most spectacular example of this is the determination of all vacuum type D solutions. This technique is not restricted to the case of vanishing cosmological constant. 

In higher dimensions, there are various notions of ``algebraically special'' that have been explored. These have been reviewed in Ref. \cite{Reall:2011ys}. The most widely used definition is that of Coley, Milsson, Pravda and Pravdova (CMPP) \cite{Coley:2004jv}. So far, the only attempt to use the CMPP ``algebraically special'' property as a tool for solving the Einstein equation is Ref.  \cite{Godazgar:2009fi} which determined all {\it axisymmetric} algebraically special solutions of the vacuum Einstein equation (with cosmological constant) where ``axisymmetric'' is defined as the existence of a $SO(D-4)$ symmetry with $S^{D-3}$ orbits. There is significant scope for extending this approach of combining the algebraically special property with some symmetry assumptions. An obvious case of interest is $U(1)^{D-2}$ symmetry.

A more systematic exploitation of the algebraically special property probably will require further development of the general theory of algebraically special solutions. The key result lacking is a higher-dimensional generalization of the Goldberg-Sachs theorem. Partial progress has been made (reviewed in Ref. \cite{Reall:2011ys}) but this is an area in which plenty remains to be done. 

An alternative definition of ``algebraically special'' involves spinors. This works only for $D=5$ (and $D=4$). This definition is independent of the CMPP classification (although the MP solutions are algebraically special according to both definitions) and therefore could lead to rather different results if exploited as a tool to solve the Einstein equation.  A first attempt at exploiting the spinorial definition of algebraically special to solve the 5D Einstein equation was made in Ref. \cite{Godazgar:2010ks}. Only the most special algebraic classes were considered. It would be very interesting to consider more general classes. 

Another property that might be exploited to solve the Einstein equation is the existence of a Killing, or Killing-Yano tensor. It is known that the wave equation is separable in a MP spacetime and this is because of the existence of a conformal Killing-Yano tensor. Furthermore, it has been shown that a certain generalization of the MP solution is the most general spacetime admitting a ``principal'' conformal Killing-Yano tensor \cite{Houri:2007xz,Krtous:2008tb}. This result shows that one can exploit the existence of such a tensor to solve the Einstein equation. In this case it leads to a known solution but maybe the assumptions here could be weakened. Is it possible to make progress if the ``principal'' condition is dropped? Does this lead to new solutions?

\subsubsection*{Classification.}

A long term goal is to solve the classification problem: classify all stationary, asymptotically flat, vacuum black hole solutions in $D>4$ dimensions. This seems far beyond current techniques.

So far, classification results for higher-dimensional black holes are restricted to $D=5$ with two rotational symmetries. Since the general case of one rotational symmetry appears intractable at present, one could consider other special cases with enhanced symmetry. For example, can one classify cohomogeneity-1 vacuum black hole solutions? Are there any such solutions other than Schwarzschild and equal angular momentum, odd $D$ Myers-Perry? What about cohomogeneity-2 solutions in general $D$? These include singly spinning MP black holes. Can such solutions be classified?

Another simplifying assumption is extremality. Any extreme black hole admits a near-horizon geometry with more symmetry than the full black hole solution \cite{Kunduri:2007vf}. Can we classify possible near-horizon geometries?(See Refs. \cite{Kunduri:2008rs,Hollands:2009ng} for results in this direction.) Can this be used to classify extreme black holes? To do this would involve understanding the global question of what restrictions asymptotic flatness imposes on a near-horizon geometry.

Although black holes with gauge fields are not within the scope of this article, it should be noted that {\it supersymmetry} is more restrictive than extremality. So perhaps one could attempt to classify supersymmetric black holes in a higher-dimensional supergravity theory. So far, this has been investigated only for the simplest 5D theory  \cite{Reall:2002bh}.

\subsubsection*{Stability.}

A long term goal is to determine the classical stability of the known black hole solutions, i.e., Myers-Perry and black rings. Where instabilities exist, the endpoint should be determined. 

As discussed above, there are now several results concerning instabilities of MP black holes. So far, studies of linearized perturbations have considered only perturbations preserving the rotational symmetry that arises from the rigidity theorem. However, for a singly spinning black hole, numerical evolution finds that the instability which appears first (i.e. at smallest $J$) breaks this symmetry. Since numerical evolution is hard, it would be interesting to study this non-rotationally symmetric instability using linearized theory. It would also be nice to know which type of instability is dominant at very large $J$. 

A target for future research would be to map out the stable regions of the MP parameter space. This could be done via a combination of linearized analysis and full numerical evolution. Although unsuccessful so far \cite{Durkee:2010qu}, perhaps one could derive a master equation describing MP perturbations, a higher-dimensional analogue of the Teukolsky equation. 

The stability of black rings against non-rotationally symmetric perturbations is an important open question that future work should address. Are there any stable vacuum black rings?

At present, the endpoint of the super-radiant instability of MP-$AdS$ black holes is unknown although it is conceivable that no regular endpoint exists \cite{Dias:2011ss}. It would be interesting to determine the time evolution of this instability numerically.

\subsubsection*{Numerical methods.}

Given that we expect higher dimensions to admit many new solutions, many without much symmetry, it is unlikely that we will succeed in constructing them all analytically. Numerically solving the Einstein equation is likely to become an important tool in finding new stationary black hole solutions. New methods have been developed recently for solving the Einstein equation to obtain stationary solutions, see Ref. \cite{Wiseman:2011by} for a review.

Of course, numerical work has its limitations: one usually must have a good idea of the properties of the solution one wishes to find: it will be hard to go ``fishing'' for new solutions; as discussed above, there may exist black hole solutions with a large number of parameters, it would be hard to explore such a parameter space numerically. Nevertheless, numerical methods can be used to confirm the existence of solutions for which the evidence is otherwise indirect, or perturbative in nature. For example, the arguments for the existence of black holes with a single rotational symmetry are based on perturbation theory. If we cannot find such solutions explicitly then numerical analysis of the Einstein equation provides the only way of confirming the existence of such solutions at the full nonlinear level. (Actually, this is not quite true: one might be able to prove existence of solutions without finding them explicitly, but such an approach may not provide much information about properties of the solutions.) Finding numerical solutions describing stationary $D>4$ black holes with a single rotational symmetry should be a goal for future work.
 
 An obvious approach to finding new black hole solutions is to try to form them in a time-dependent simulation of gravitational collapse. However, this approach probably will only find classically stable solutions and many higher-dimensional solutions are expected to be unstable. (This does not make them uninteresting: the time-scale for the instability might be long e.g. compared to the time-scale of Hawking evaporation.) On the other hand, there is no reason that new stable solutions should not exist, particularly in the less well understood case of asymptotically anti-de Sitter spacetimes. 
 
 The recent discovery of the absence of a threshold for black hole formation \cite{Bizon:2011gg} emphasizes how little is understood about time dependent processes in $AdS$. Numerical simulations will be invaluable in developing our intuition for what is possible. It will be interesting to see how the results of Ref. \cite{Bizon:2011gg} are modified if one relaxes the assumption of spherical symmetry. 

The case of asymptotically $AdS$ spacetimes is particularly interesting because of the range of possible boundary conditions that one can consider. For example, there has been recent progress in constructing solutions in which the conformal boundary is the Schwarzschild spacetime \cite{Figueras:2011va}, which can be used to study the behaviour of strongly coupled field theory in a black hole background. This idea has been discussed more generally in Ref. \cite{Hubeny:2009ru}, which argues for the existence of different types of ``funnel'' and ``droplet'' solutions when the boundary metric describes a black hole. Future numerical work should investigate such solutions. It will also be interesting to construct numerically solutions which describe an $AdS$ black hole localized on an internal space e.g. on the $S^5$ of $AdS_5 \times S^5$ (see \cite{Hubeny:2002xn} for a discussion of the instability of Schwarzschild-$AdS_5\times S^5$ and its implications for the dual gauge theory).

\subsubsection*{Approximate methods.}

The equations of the blackfold approach described above, which allow to
construct approximate black hole solutions, have been solved only in the
simplest cases with a high degree of symmetry that allows to solve the
equations algebraically. This, however, is far from being a systematic
and exhaustive study and has uncovered only a very small part
of the solutions that presumably exist in $D\geq 6$. 
For instance, it would be important to study
solutions with a lower degree of symmetry, since they will play a role
in connecting different black hole phases, as well as providing new
classes of black holes with less symmetry than the Cartan subgroup of
$SO(D-1)$. Novel horizon topologies may also be discovered. 

Instead of a
case-by-case study of specific ansatze for the solutions, a more
systematic approach to the blackfold equations would be highly desirable. 
This
approach should give very valuable results towards a classification of
black holes, or at least towards a useful characterization of them.

The blackfold methods can also be usefully applied to the study of the
stability properties of black holes whenever their horizons possess two
separate length scales. The connection it makes to hydrodynamics has
made it possible to simplify the study of the GL
instability of black branes and possibly will yield a better
understanding of the non-linear evolution of this instability towards
its endpoint. Having an analytical tool to study this problem seems very
valuable since so far it has only been tackled via massive numerical
calculations with supercomputers.

%% file: Transplanckian_scattering.tex
\section{Trans-Planckian scattering} \label{sec:tp}
\begin{center}
Coordinator: Seongchan Park
\end{center}

The ultimate aim of physics is to discover the fundamental laws of nature. According to the uncertainty principle, $\Delta x \geq \hbar /\Delta p$, higher energies are needed to probe smaller distances.  Ultimately, however, when the energies involved in physical processes exceed the Planck energy we enter the profoundly mysterious trans-Planckian regime.  
In trans-Planckian processes, gravity dominates the scattering at increasing distances~\cite{'tHooft:1987rb,Giddings:2010pp}, and it has been argued that it prevents probing  distances shorter than the Planck length~\cite{Banks:1999gd,Giddings:2001bu}, 
\be
\ell_D\equiv \left(\frac{G_D\hbar}{c^3}\right)^{\frac{1}{D-2}}\ .
\ee
 Specifically, in the regime with centre of mass energy $\sqrt{s} \gg M_D$, where the Planck mass in $D$-dimensions is $M_D=\left(\frac{\hbar^{D-3}c^{5-D}}{G_D}\right)^{\frac{1}{D-2}}$, the Schwarzschild-Tangherlini black hole radius ($R_s$) exceeds the de Broglie wavelength,
\be
R_s=\left(\frac{16\pi}{(D-2)\Omega_{D-2}}\right)^{\frac{1}{D-3}} \left(\frac{G_D \sqrt{s}}{c^4}\right)^{\frac{1}{D-3}} \gg \lambda =\frac{4\pi \hbar}{\sqrt{s}}\ 
\ee
and the scattering range grows as a power of energy.
 Here, general features of the dynamics are expected to be well-approximated by semiclassical scattering and/or horizon formation.

\begin{figure}[h]
\centering
\includegraphics[width=0.75\textwidth]{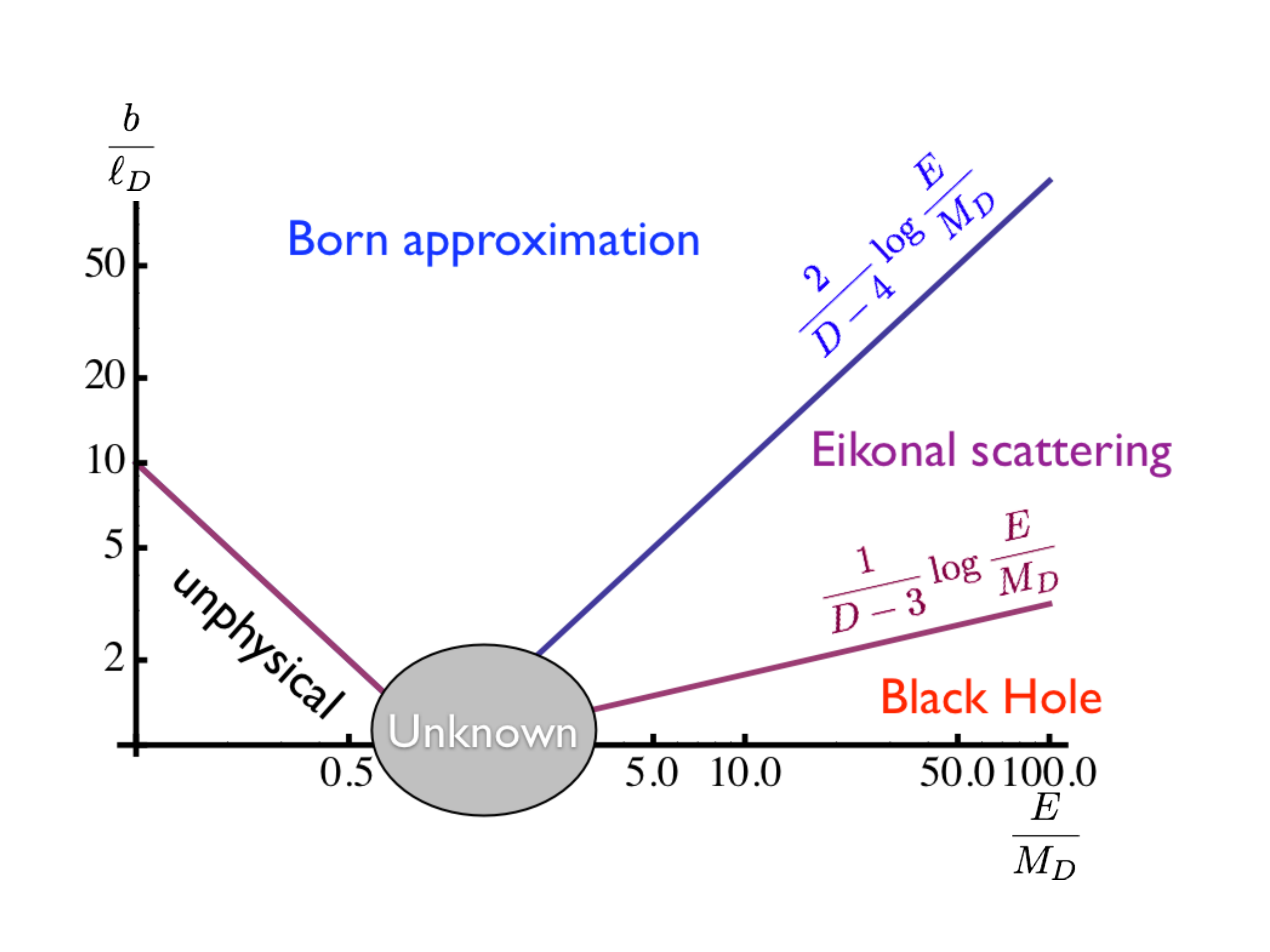}
\caption{A proposed ``phase diagram'' of different regimes for gravitational scattering (from \cite{Giddings:2011xs}).}
\label{Fig:phase}
\end{figure}

One can characterize the scattering in terms of the center-of-mass energy $E=\sqrt{s}$ and impact parameter $b = J/\sqrt{s}$, where $J$ is the angular momentum.  
A proposed `phase diagram' describing the relevant physics in different regions of the $(E,b)$ plane is summarized  in Fig. \ref{Fig:phase} (see \cite{Giddings:2011xs}). The left lower part is unphysical as $\Delta x < \hbar/\Delta p$. The gray part near the Planck scale $\sqrt{s}\sim M_D$ is largely unknown and non-linear quantum gravity effects dominate so that we need the full theory of quantum gravity to describe physics in this domain. On the other hand, the large impact parameter regime appears well-approximated by the Born or eikonal approximations, as long as the 
scattering angle is `small', $\theta \sim \left(R_s/b\right)^{D-3}\ll 1$. Lowering the impact parameter further to the value of Schwarzschild-Tangherlini radius,
%
gravity becomes strong and highly non-linear, and black hole formation is expected even though a full quantum description of this process remains unknown. Specifically, in the classical theory high-energy particles are well approximated by colliding Aichelburg-Sexl solutions~\cite{Aichelburg:1970dh}, and in such a collision, at impact parameters $b\lesssim R_s(\sqrt s)$, one can show that a trapped surface forms~\cite{Penrose,Eardley:2002re,Yoshino:2002tx}.  
(One can supply additional arguments that quantum effects do not modify these statements~\cite{Giddings:2004xy}.)
Simulations in numerical relativity have confirmed this prediction of black hole formation in high-energy collisions \cite{Choptuik:2009ww,Sperhake:2008ga,Shibata:2008rq,Sperhake:2009jz}. 

Once a black hole is formed, it decays through Hawking radiation.  This process would enable the LHC to probe black hole signatures  in TeV gravity models with extra dimensions~\cite{Giddings:2001bu,Dimopoulos:2001hw} (For reviews, see for instance Refs.~\cite{Giddings:2007nr,Kanti:2004nr,Cavaglia:2002si,Landsberg:2006mm}).

In this section, we review the current status of trans-Planckian physics focusing on the LHC search for black holes and other trans-Planckian signatures.
%

\subsection{Is Trans-Planckian physics testable in the near future?: the experimental status of TeV gravity models}

The critical question for experimental observation of trans-Planckian  physics is the size of the fundamental Planck scale.
In extra-dimensional models with large and/or highly warped extra dimensions, the scale of gravity can be as low as $M_D={\cal O}(1)$ TeV, eliminating a large hierarchy between this scale and the electroweak scale.   
The LHC has begun to exclude the low end of the parameter space for TeV gravity models by searching for signatures in various channels.  What is the current status of the exclusion? 
What are the future expectations? 
\vspace{1cm}

\subsubsection*{ATLAS searches.}
At the time of writing, the most recent results from ATLAS come from an integrated luminosity of around $1\mathrm{fb}^{-1}$, acquired during the first part of 2011. These results will be updated with around 5 pb$^{-1}$ available in early 2012. ATLAS has focused its searches on channels with leptons, because of the expected democratic nature of the gravitational coupling. This means that one would expect particle production to be in proportion to the number of states available (with appropriate gray-body factors), and not related to the gauge quantum numbers of the states. However, it is possible to construct models, such as split branes, where this is not the case. 

In \cite{ATLAS2011LepJet} a search for microscopic black holes and string ball states was performed, using final states with three or more high transverse momentum objects, at least one of which was required to be an electron or muon. No deviations were observed from the standard model expectations, which were estimated using a combination of data-driven and Monte Carlo based techniques. The dominant background sources come from vector boson production, either directly or from top decays. A smaller background arises from QCD events which contain a fake lepton. The events were studied as a function of the scalar sum of the transverse momenta of the final state particles ($\Sigma p_T$). In the highest bin, with $\Sigma p_T>1500$~GeV, there were 8 (2) electron (muon) events observed, with standard model expectations of $10.2\pm1.4\pm2.6$  ($2.8\pm0.5\pm1.1$), respectively.

In \cite{Collaboration:2011bw} events are selected containing two muons of the same charge. This channel is expected to have low Standard Model backgrounds while retaining good signal acceptance. Isolated muons (i.e. muons with very little activity around them in the detector) can be produced directly from the black hole or from the decay of heavy particles such as W or Z bosons. Muons from the semi-leptonic decays of heavy-flavour hadrons produced from the black hole can have several other particles nearby and can therefore be non-isolated. In order to maintain optimal acceptance for a possible signal, only one of the muons is required to be isolated in this analysis, thereby typically increasing the acceptance in the signal region by 50\%.
The decay of the black hole to multiple high-$p_T$ objects is used to divide the observed events into background-rich and potentially signal-rich regions. This is done by using the number of high-$p_T$ charged particle tracks as the criterion to assign events to each region. Black hole events typically have a high number of tracks per event ($N_{trk}$), while Standard Model processes have sharply falling track multiplicity distributions. In the background-rich region, where only small signal contributions are expected, data and Monte Carlo simulations are used to estimate the number of events after
selections. This background estimate is validated by comparing to data. The expected number of events from Standard Model processes in the signal-rich region is then compared with the measured number, and a constraint on the contribution from black hole decays is inferred. Good agreement is observed between the measured distributions and the background expectations. No excess over the Standard Model predictions is observed in the data.

Although it is clear that there is no anomalous signal in the data, setting limits on black hole production is problematic. The LHC experiments are searching at the limit of its current energy range, and hence, by definition, cannot explore well beyond the current limit on the bulk Planck mass. This means that predictions for the rate of black hole production using semi-classical approximations are not valid: a full theory of quantum gravity is required. 

The experiment tackles this issue in two ways. Firstly, using the number of events observed in data, and the background expectations, upper limits are set on $\sigma \times BR \times A$, where $\sigma$ is the cross section, BR the branching ratio to the signal channel, and A the acceptance of non-Standard Model contributions in this final state in the signal region. These limits are reasonably model-independent. However, to make use of these limits, theorists need to know the relevant value of A for their model scenario. Representative values of A are therefore published by the experiments. The acceptance tends to be high for the very high mass states from black hole decay, and so this approach is useful; however, such limits must be used with caution when applied to extreme scenarios, such as two-body decays.

The second approach is to set limits on benchmark scenarios. This is done in a plane of $M_D$ vs $M_{TH}$, where $M_D$ is the bulk Planck mass. For the purpose of limit setting, it is assumed that black hole production only occurs above $M_{TH}$, which should be well above $M_D$ for the semiclassical calculations used in the event generators to be valid. This approach produces 90\% confidence limits up to $M_D \approx 1.5$ TeV depending on the model used. However, there is a strong dependence on the modeling of the final remnant decay, which is a pure quantum gravity effect, and can dominate the final state in this mass range.

\subsubsection*{CMS Searches}

The CMS experiment has pioneered accelerator searches for black holes in 2011 and published a paper~\cite{CMSBH} based on data collected in the first 7 TeV running period of the LHC (March-November, 2010), corresponding to an integrated luminosity of 36 pb$^{-1}$. Despite the relatively small statistics, the sensitivity of the search was high enough to largely disfavor the possibility of black hole production at a 7 TeV center-of-mass energy. The analysis conducted by the CMS collaboration was done in the inclusive final state, thus maximizing the sensitivity to black-hole production and decay. Semi-classical black holes are expected to evaporate in a large ($\sim 10$) number of energetic particles, emitted nearly isotropically, with the major fraction of the emitted particles being quarks and gluons, which are detected as jets in the CMS detector. Quantum effects and gray-body factors may change the relative fraction of emitted quarks and gluons, but generally it is expected that these particles appear most often even in decays of quantum black holes, due to a large number of (color) degrees of freedom that quarks and gluons possess, compared to the other standard model particles (e.g., photons, leptons, neutrinos, and gauge bosons).

The discriminating variable between the signal and the dominating QCD multijet background used in the search was the scalar sum of transverse energies\footnote{Transverse energy of a particle, $E_T$, is defined as the energy of the particle $E$ times the sine of the polar angle of the particle direction with respect to the counterclockwise proton beam.} of all particles in the event, for which transverse energy exceeds 50 GeV. This variable, $S_T$, was further corrected for any missing transverse energy in the event by adding the missing transverse energy to the $S_T$ variable, if the former exceeds 50 GeV. The choice of $S_T$ as the discriminating variable is extremely robust and rather insensitive to the particle content in the process of black-hole evaporation, as well to the details of the final, sub-Planckian evaporation phase. The addition of the missing transverse energy to the definition of $S_T$ further ensures high sensitivity of the search for the case of stable non-interacting remnant with the mass of order of the fundamental Planck scale, which may be produced in the terminal stage of the evaporation process.

The main challenge of the search is to describe the inclusive multijet background in a robust way, as the black-hole signal corresponds to a broad enhancement in $S_T$ distribution at high end, rather than a narrow peak. Since the black-hole signal is expected to correspond to high multiplicity of final-state particles, one has to reliably describe the background for large jet multiplicities, which is quite challenging theoretically, as higher-order calculations to fully describe multijet production simply do not exist. Thus, one can not rely on the Monte Carlo simulations to reproduce the $S_T$ spectrum correctly. 

\begin{figure}[hbt]
\begin{center}
\includegraphics[width=0.48\textwidth]{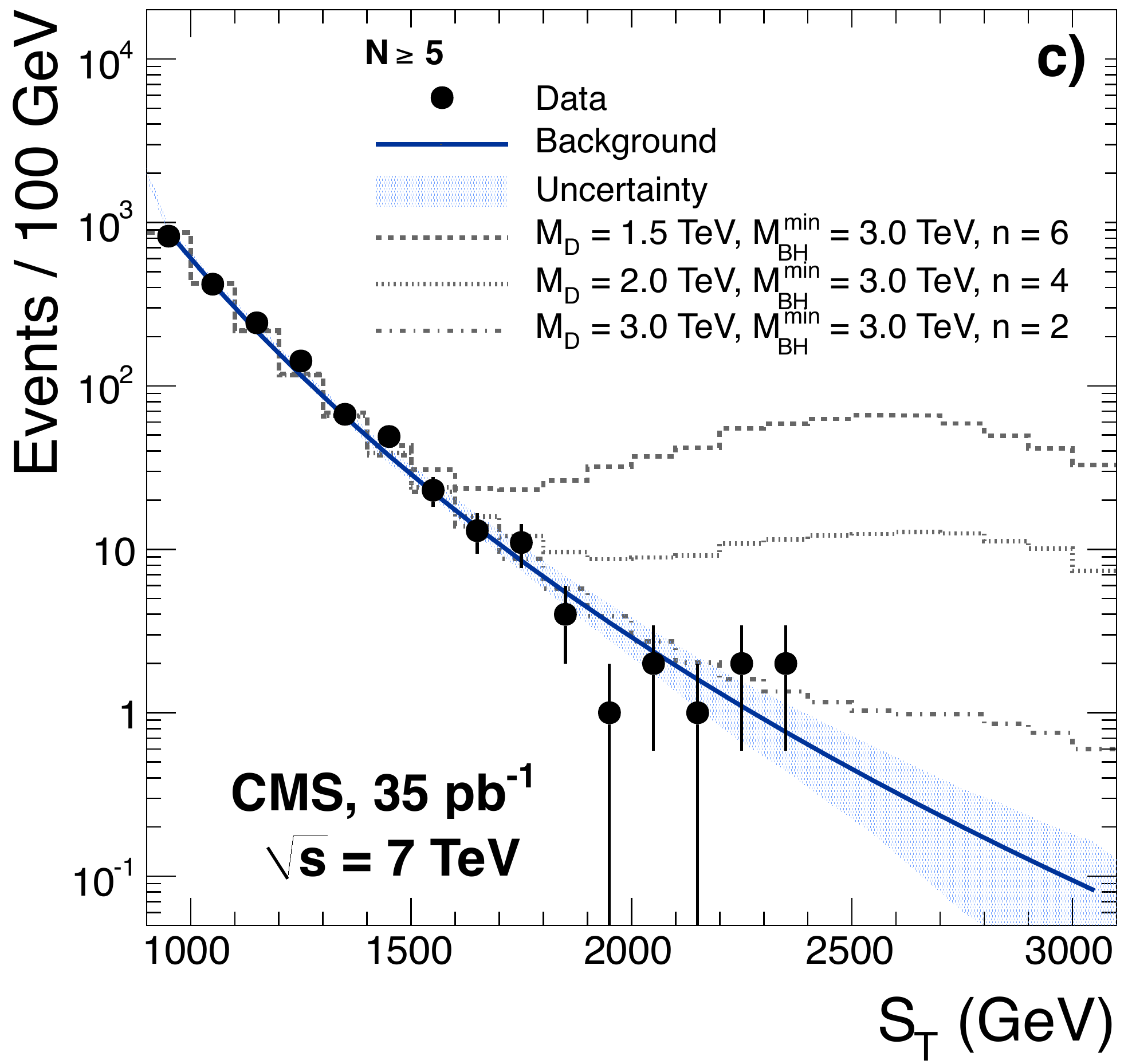}
\includegraphics[width=0.48\textwidth]{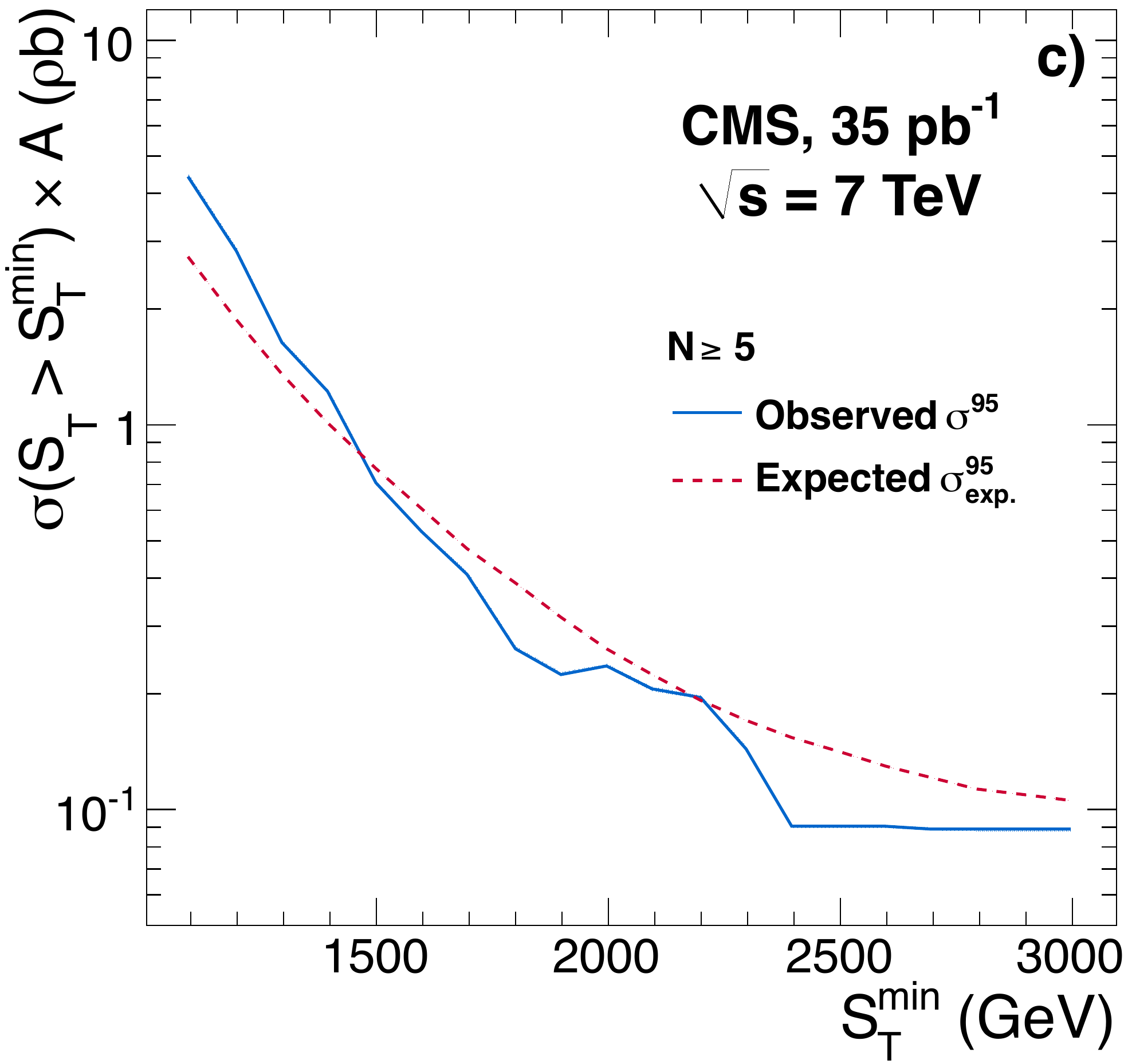}
\end{center}
\caption{Left: predicted QCD multijet background with its uncertainties (the shaded band), data, and several reference black-hole signal benchmarks, as a function of $S_T$ in the final state with the multiplicity of 5 or more particles. Right: model-independent upper limits at 95\% confidence level on a cross section of a new physics signal decaying in the final state with 5 or more particles, as a function of the minimum $S_T$ requirement. From \protect{\cite{CMSBH}}.}
\label{fig:CMS1}
\end{figure}

The CMS collaboration developed and used a novel method of predicting the QCD background directly from collision data to overcome this problem. It has been noted empirically, first via Monte Carlo based studies, and then from the analysis of data at low jet multiplicities that the shape of the $S_T$ distribution for the dominant QCD multijet background does not depend on the multiplicity of the final state, above the turn-on threshold. This observation, motivated by the way parton shower is developed via nearly collinear emission, which conserves $S_T$, allows one to predict $S_T$ spectrum of a multijet final state using low-multiplicity QCD events, e.g. dijets or three-jet events. This provides a powerful method of predicting the main background for black-hole production by taking the $S_T$ shape from the diet events, for which signal contamination is expected to be negligible, and normalizing it to the observed spectrum at high multiplicities at the low end of the $S_T$ distribution, where signal contamination is also negligible even for large multiplicities of the final-state objects. The results are shown in Fig.~\ref{fig:CMS1} (left).

\begin{figure}[hbt]
  \begin{center}
       \includegraphics[width=0.6\textwidth]{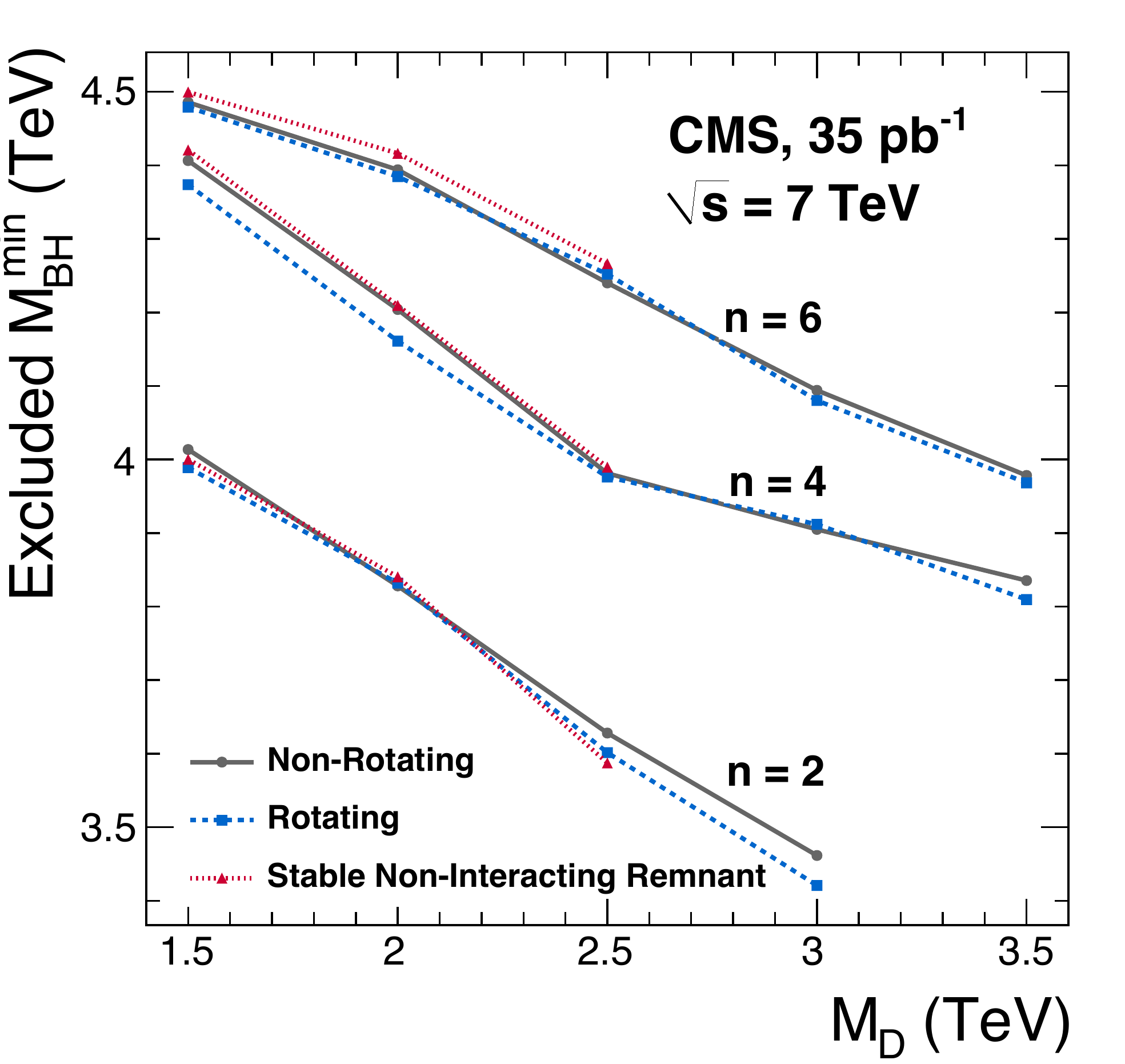}
  \end{center}
\caption{Limits on the minimum black-hole mass as a function of the fundamental Planck scale for a few semi-classical benchmark models with and without black hole rotation and non-evaporating remnant. Note that the semi-classical approximation used in setting these limits is not expected to hold for black hole masses so close to the Planck scale, so this plot should be considered as illustration only. From \protect{\cite{CMSBH}}.}
\label{fig:CMS2}
\end{figure}

The CMS data at high final-state multiplicities is well fit by the background shape obtained from the diet events, no excess characteristic of a black-hole production is seen in the data. This lack of an apparent signal can be interpreted in a model-independent way by providing  limit on a cross section for any new physics signal for $S_T$ values above a certain cutoff, for any given inclusive final state multiplicity. An example of such a limit is shown in Fig.~\ref{fig:CMS1} (right), for the final-state multiplicity $N \ge 5$. For signals corresponding to large values of $S_T$ (above 2 TeV or so) the cross section limit reaches $\sim 100$ fb. These limits can be compared with production cross section for the black holes in a variety of models and used to set limits on the minimum black-hole masses ($M_{\rm BH}$) that can be produced in these models. In the original publication~\cite{CMSBH}, the CMS collaboration used just a few simplest semi-classical benchmark models that are expected to break down at the values of black-hole masses near the exclusion, given that they are of the same order as the probed values of the fundamental Planck scale. Despite these obvious limitations, the benchmark models probed in this search show that the results are rather insensitive to the details of the black-hole decay and that black-hole masses as high as 4.5 TeV are probed in this analysis. The negative results of this search largely exclude the possibility of observing black-hole production at the 7 TeV LHC. An updated analysis using full 2011 statistics is expected to be published soon.

\subsubsection*{Theoretical issues.}

In order to observe strong-gravitational scattering, approximated as classical black hole formation, scattering energies must significantly exceed the Planck energy.  There are two reasons for this:  1) an object must have a mass significantly exceeding $M_D$ in order to be well-approximated as a semiclassical black hole; the expansion parameter justifying the semiclassical approximation is $M_D/E$ and 2) not all of the energy of colliding partons goes into black hole formation; some escapes in radiation~\cite{D'Eath:1992hb,Giddings:2001bu,Eardley:2002re}.   

It is thus important to determine the threshold for onset of black hole formation, for a given Planck mass.  There are different aspects of this problem.  Classically, an important question is to determine the inelasticity, the amount of energy lost to radiation as compared to the mass of the resulting black hole.  The trapped surface constructions \cite{Penrose,Eardley:2002re,Yoshino:2002tx} places lower bounds on the mass of the black hole; these have been modestly improved in \cite{Yoshino:2005hi}. One can also estimate the radiated energy perturbatively
\cite{D'Eath:1992hb,D'Eath:1992hd,D'Eath:1992qu,Herdeiro:2011ck,Cardoso:2002ay,Berti:2010ce,Berti:2010gx}. Here, numerical relativity calculations could ultimately provide more definitive answers, by computing the amount of radiated energy, in higher dimensional gravity, and as a function of impact parameter.

Another aspect of the problem is to determine where quantum effects lead to large deviations from semiclassical black hole behavior.  This requires greater knowledge of quantum gravity. One possible criterion for onset of black hole behavior is the onset of thermodynamic behavior, characterized by the entropy \cite{Giddings:2001bu}.

\subsection{How can we improve our understanding of trans-Planckian Scattering ?}
There are a number of important questions on the subject of trans-Planckian scattering.
As noted above, we would like to determine classical characteristics of high-energy gravitational collisions. In particular, we would like to find the radiated energy and black hole mass as a function of collision energy and impact parameter.  In addition to giving the inelasticity factors, this would also provide information about the cross section for black hole formation.  It would also be interesting to determine further characteristics of the radiation, and the evolution of the black hole as it goes through the ``balding" phase~\cite{Giddings:2001bu}, asymptoting to a Myers-Perry \cite{Myers:1986un} black hole.  Also of interest is determining the classical evolution (e.g. radiation details) in the larger impact parameter regime, where a black hole does not form.  

A second class of questions regards corrections to the classical analysis.  Specifically, given a classical description of scattering in the eikonal and black hole regimes, one would like to understand the size and nature of quantum corrections.  In the eikonal regime $\theta\ll1$, one would for example like to check what effect quantum corrections have on the classical scattering plus radiation, and check the expectation that the basic features of the scattering are largely determined by classical physics.

In the black hole regime, we know there are important quantum corrections to the classical analysis.  These include Hawking radiation.  However, so far we lack any quantum description of the full evolution of the black hole, and in fact encounter a conflict of basic physical principles in attempting to describe such evolution.  This conflict (for some more discussion see \cite{Giddings:2011xs} and references therein) has been called the ``information paradox" or ``unitarity crisis," and seems to indicate the need for fundamental revision of the foundations of physics.  Beyond theoretical study of this regime, one can anticipate that experimental data on transplanckian scattering, should it be found, could provide further clues.

\subsubsection*{Interplay between perturbative and numerical methods.}

Despite being a well defined problem the computation of the exact solution for the collision of two ultra-relativistic particles in general relativity is out of reach and approximation techniques have to be used. Then, obtaining estimates from different techniques becomes fundamental for cross-checking. Let us discuss three such techniques and compare their results.

The oldest method consists in modeling the gravitational field of the colliding particles by Aichelburg-Sexl (AS) shock waves \cite{Aichelburg:1970dh} and compute the apparent horizon that forms, as first done in four dimensions by Penrose \cite{Penrose}. 
From the apparent horizon construction a lower bound for the energy loss into gravitational radiation and estimate of the threshold impact parameter for black hole formation in $D$-dimensional collisions can be obtained \cite{Eardley:2002re}. This method does not require the knowledge of the geometry in the future light cone of the collision and can only provide bounds. Its main advantage is its technical and conceptual simplicity, although the critical impact parameter estimates in $D>4$  require numerical solutions \cite{Yoshino:2002tx}. This method suggests that the amount of radiation loss increases (as a ratio of $\sqrt{s}$) with $D$. An improvement of the method, for determining the cross section,  that considers an apparent horizon \textit{on} the future light-cone (but not inside) was given in \cite{Yoshino:2005hi}.

The second method, which may be regarded as a refinement of the first one, also uses the superposition of two AS shocks, but attempts to compute the geometry \textit{in} the future light cone of the collision by a perturbative expansion. In four dimensions the method was carried out to second order in perturbation theory \cite{D'Eath:1992hb,D'Eath:1992hd,D'Eath:1992qu}, in the case of a head-on collision, yielding $0.164\sqrt{s}$ for the energy loss. The case with impact parameter is technically harder, due to the loss of symmetry, and has not been fully discussed. The generalisation of this method to higher $D$ head-on collisions was done recently, in first order perturbation theory \cite{Herdeiro:2011ck}. Interestingly, the trend exhibited for the energy loss is the same as that obtained from the first method and suggests that the two methods converge as $D\rightarrow \infty$.

The third method is to model the colliding particles as black holes 
and to perform high energy collisions using numerical relativity techniques. The use of black holes is, in principle, only for simplicity (i.e. one uses the vacuum Einstein equations). Using any other lumps of energy (such as boson stars) should yield similar results in the trans-Planckian regime, since the process should be dominated by the energy of the objects; in other words, the individual phase space of the colliding objects is irrelevant as compared to the phase space of the entire collision process, and the detailed structure of the objects is hidden behind their mutual event horizon. It would be extremely interesting to confirm this expectation; in four dimensions, since high energy head-on collisions of both black holes \cite{Sperhake:2008ga} and boson stars \cite{Choptuik:2009ww} have been performed, such comparison is within reach. The high energy collisions performed so far have reached $\gamma\sim 3$ \cite{Sperhake:2008ga}; these authors extrapolated to $\gamma\rightarrow \infty$ using the numerical points and a fit based on the zero frequency limit\footnote{The error bars are dominated by the numerical errors, rather than the specific fit.}, yielding, in the ultra-relativistic regime $(0.14\pm 0.03)\sqrt{s}$ for the energy loss. Remarkably, this overlaps the second order perturbation theory result described above, which may be interpreted as lending credibility to both methods. Thus, two important directions are to obtain both the perturbative computation to second order in higher $D$ and the corresponding numerical simulations. Some important steps concerning the latter have been given already \cite{Yoshino:2009xp,Zilhao:2010sr,Witek:2010xi,Witek:2010az,Zilhao:2011yc}, but it seems harder to reach the same $\gamma$'s in higher dimensions, due to stability, but also perhaps fundamental \cite{Okawa:2011fv}, problems.

\subsection{Black hole search at the LHC: future improvements}

There are various theoretical developments needed to refine black hole searches in high-energy collisions.  As noted above, it is important to better determine black hole mass as a function of energy and impact parameter, which also supplies cross section information, as well as to characterize the classical radiation.

To improve study of black hole signatures, one needs further information about black hole decay.  In particular, the full evolution of a black hole through the ``spindown'' and ``Schwarzschild'' phases \cite{Giddings:2001bu}, ending with the ``Planck phase'' $M\sim M_D$, where the semiclassical approximation breaks down, has not been determined.  One would like to determine from this the spectra of decay products, as well as other features such as angular distributions, indicative of black hole spin.  First estimates of these basic features for a spinning black hole were given in \cite{Giddings:2001bu}, and there has been a lot of work on the important problem of refining calculations of gray-body factors \cite{Ida:2002ez,Ida:2005zi,Ida:2005ax,Ida:2006tf}, \cite{Harris:2005jx,Duffy:2005ns,Casals:2005sa,Casals:2006xp,Casals:2008pq, Sampaio:2009tp, Sampaio:2009ra, Rogatko:2009jp, Kanti:2010mk,Herdeiro:2011uu,Cardoso:2005vb,Cornell:2005ux,Kanti:2009sn}.  However, we still lack a calculation of the gray-body factors for graviton emission from spinning black holes, and so a complete picture of black hole evolution is lacking (see \cite{Ida:2006tf} for the evolution in the spin-down and Schwarzschild phases).  Such a calculation in particular is important to determine the amount of radiation in visible particles.

Finally, in addition to the questions associated with the unitarity crisis (described above) and Planck phase, there are questions of the dependence of black hole decay on the detailed microphysical model of TeV-scale gravity.  Such details could contribute additional signatures.  For this reason, further study of viable models for TeV-scale gravity would be helpful.


The data samples accumulated during 2011 are likely to provide the strongest limits obtainable on black hole production for the LHC at a centre of mass energy of 7 TeV. The results are already constrained by the lack of theoretical understanding of the final stages of black hole decay. Furthermore, the fundamental Planck scale $M_D$ has been constrained over the years, up to the present LHC data. The main sources of bounds come from: i) deviations from Newtonian gravity in torsion balance experiments~\cite{kapner:021101,tu:201101}; ii) collider searches for Kaluza-Klein graviton production (monojet or photon with missing energy)~\cite{Aaltonen:2008hh,Aad:2011xw,Chatrchyan:2011nd} and KK graviton mediated dilepton or diphoton production~\cite{LEPexotica_graviton,Abazov:2008as,Abazov:2005tk,ATLAS:2011ab,Chatrchyan:2011fq}; iii) astrophysical or cosmological KK graviton production in supernovae and in the early universe~\cite{Hannestad:2001xi,Hannestad:2001nq}. The most stringent laboratory bounds from current LHC searches indicate $M_D \gtrsim 2.5-3.5~\mathrm{TeV}$ for $D>6$ whereas observational bounds from astrophysics and cosmology only allow for such small $M_D$ for $D>7$ (however there are many model dependent uncertainties on the latter). Because of the theoretical uncertainties and the limits on $M_D$ indicated by these searches, it is important to perform a wide-ranging search in the data, using all available channels to look for TeV gravity effects. It would be dangerous, for example, to use limits from dijet production to infer an absence of signal in a leptonic channel, since a TeV gravity model would be needed to link the observations together. The experiments have already looked at dijet, multijet, single lepton and dilepton data, and this approach should be continued. Once the full 2011 dataset has been analysed, further progress will probably have to await an increase in the LHC beam energy. 

On the theoretical front, it would be useful to try to constrain the range of allowed scenarios near the bulk Planck mass. The cross section is predicted to rise by many orders of magnitude, at a very rapid rate, in this region. From past experience with the onset of new physics, such as that observed in low energy hadron scattering near the QCD scale, one would expect resonant behavior near threshold, settling to the semi-classical prediction as the energy increases. Such behavior could greatly enhance the sensitivity of the experiments, since the peak cross sections could be very large. It would be useful, if it were possible, to make generic predictions for the maximum allowed cross section for black hole production, perhaps based on unitarity considerations, or even on the maximum allowed rate of change of the cross section near threshold. Such predictions could then be used instead of sampling parameter space in regions where the approximations required are known to be violated.


Finally, studies of the possible black hole solutions of higher-dimensional gravity over the last decade have revealed a vast number of possible solutions besides the Myers-Perry family of black holes (see Section \ref{BHs}). These include five-dimensional black rings and black Saturns (which are known in exact explicit form) as well as a large number of other black holes in six or more dimensions with more complicated topologies (which have been constructed using approximate techniques). The potential formation and detection of one of these black holes would provide a new window into the study of higher-dimensional gravity. However, many of these solutions are expected to be dynamically unstable when their angular momentum is sufficiently large and therefore would not appear as states after the balding phase (although they might play a role in the approach to this phase). For moderate angular momenta, currently it is not yet known whether a black ring (arguably the simplest and most important among the new kinds of black holes) may be dynamically stable and thus have observational signatures in a collision. It is expected that progress in this stability problem will be achieved in the near future via numerical studies. Before that, it seems premature to make any predictions about the possibility of observing them in collider experiments.



%% file: Holo.tex
\section{Strong gravity and High Energy Physics} \label{sec:holography}
\begin{center}
Coordinator: Paul~M.~Chesler
\end{center}
\subsection{Dynamics in holographic quantum field theories from numerical relativity}

The study of real-time dynamics in quantum field theory (QFT) is a theoretically diverse topic, with 
available tools ranging from perturbative quantum field theory and kinetic theory to string theory and gauge/gravity duality.
It's also a challenging topic to study.  Prior to the discovery of gauge/gravity duality, the only theoretically controlled
regime of real-time dynamics in QFT was that of asymptotically weak coupling, where QFTs often admit an effective kinetic theory description
in terms of weakly interacting quasiparticles.  In this regime one can systematically study real time dynamics via perturbative expansions.  However,
the domain of utility of real-time perturbative expansions is more often than not quite limited, with extrapolations to $O(1)$ couplings converging 
poorly.

One interesting phenomena where strongly coupled dynamics appears to be relevant is that of heavy ion collisions
at the Relativistic Heavy Ion Collider (RHIC) and the Large Hadron Collider (LHC).  There, collisions are believed to produce a 
strongly coupled quark-gluon plasma (QGP), which behaves as a nearly ideal liquid \cite{Shuryak:2004cy}.  During the initial stages of a collision, when the QGP is 
produced, the system is surely far-from-equilbrium.  However, the success of near-ideal hydrodynamic models of heavy ion collisions 
suggests that the time required for the far-from-equilibrium initial state to thermalize 
may be as short as 1 fm/c, the time it takes for light to traverse the diameter of a proton \cite{Heinz:2004pj}.

Quantum Chromodynamics (QCD) is the accepted theory of the strong interactions and therefore should describe the dynamics of heavy ion collisions.
Understanding the dynamics of the QGP produced in heavy ion collision --- from small viscosities to short thermalization 
times --- from QCD has been a challenge.  This isn't to say that QCD is an incorrect description of nature, 
but rather that theorists simply do not know how to do controlled calculations in QCD when the coupling is large.

Holography, or gauge/gravity duality, has emerged as a powerful tool to study real-time dynamics in strongly coupled QFTs from first principles calculations \cite{Maldacena:1997re}.  
The utility of gauge/gravity duality
lies in the fact that it maps the dynamics of some strongly coupled QFTs (\textit{i.e.} holographic QFTs) in $D$ spacetime dimensions onto the dynamics of semiclassical 
gravity in $D+1$ spacetime dimensions.  From a utilitarian perspective this means strongly coupled QFT dynamics 
can be mapped onto classical PDEs in $D+1$ spacetime dimensions, which can be solved numerically if needed.

Holography is a unique tool.  There is no other tool available which provides controlled access to dynamics in strongly coupled 
QFTs.  However, it is also a tool of limited applicability  \cite{Mateos:2011bs}.  Currently, there are no known theories of nature with dual gravitational descriptions.
However, this has not deterred the construction of holographic toy models of QCD or other QFTs.  While holography does not 
provide systematic access to strongly coupled QCD, it does provide systematic access to a regime of QFT never accessible before
and therefore should be explored to the fullest possible extent.  Moreover, holographic toy models of QCD can provide valuable qualitative insight 
into strongly coupled dynamics in QCD (for a useful review see \cite{CasalderreySolana:2011us}).  
Perhaps the most celebrated example is that of the shear viscosity in strongly coupled holographic QGP.
All QGPs with holographic descriptions have the same shear viscosity to entropy density ratio $\frac{\eta}{s} = \frac{1}{4 \pi}$ \cite{Kovtun:2004de}.  
Such a small viscosity is a hallmark of a strongly coupled QFT.
Indeed, recent models of heavy ion collisions suggest that the viscosity of the QGP produced is within a factor of two 
of $\frac{\eta}{s} = \frac{1}{4 \pi}$ \cite{Luzum:2008cw}.   No other systematic calculation of the viscosity 
has yielded values close to this.

Much can be learned about holographic QFTs via analytical techniques.  Indeed, many other transport coefficients than the shear viscosity 
can be computed analytically \cite{Bhattacharyya:2008jc, Baier:2007ix}.  However, there are many dynamical processes where one must resort to numerical techniques.  For example,
the creation of a QGP via collisions of sheets of matter in $3+1$ spacetime dimensions is dual to the creation of a black hole via the collision of gravitational waves in $4+1$ dimensions.
To study such a complicated process one must have a numerical relativity toolkit suitable for spacetimes relevant to holography.  The development of such a toolkit 
would be a valuable addition to the quantum field theorist's toolbox.  However, as we discuss below in Sec.~\ref{challenge}, stable algorithms for numerical relativity in holographic spacetimes 
have been difficult to construct.  As a consequence of this, very little progress has been made in the application of numerical relativity to holography and strongly coupled 
dynamics in QFT.


\subsection{The challenge of numerical relativity in asymptotically $AdS$ geometries}
\label{challenge}

The simplest and most widely studied example of holography is the $AdS$/CFT correspondence, which maps the dynamics of non-Abelian conformal field theories (CFT)
in $D$ spacetime dimensions onto semiclassical gravity in asymptotically $AdS_{D+1}$ spacetime.  In particular, the $AdS$ metric
encodes the expectation value of the stress $\langle T_{\mu \nu} \rangle$ in the dual CFT \cite{deHaro:2000xn} and metric correlation functions (which 
satisfy linearized Einstein equations) encode stress correlation functions in the dual CFT \cite{CaronHuot:2011dr}.  
For concreteness we shall consider the 
case of $D = 3+1$.  In Fefferman-Graham coordinates, the metric 
of asymptotically $AdS_5$ takes the form
\begin{equation}
\label{metric}
ds^2 = r^2 g_{\mu \nu}(x,r)dx^\mu dx^\nu+ \frac{dr^2}{r^2},
\end{equation}
where $x^\mu$ are the $3+1$ spacetime directions of the dual CFT and $r$ is the radial direction of the $AdS$ geometry.
The boundary of the $AdS$ geometry, which is where the dual CFT can be thought of as living, is at $r = \infty$.  

As is evident from the metric (\ref{metric}), a lightlike signal from some finite $r$ can reach $r = \infty$ in a finite amount of time.
As a consequence of this, boundary conditions must be imposed on the metric at the boundary.
To shed light on what the requisite boundary conditions should be, it is useful to solve 
Einstein's equations with a series expansion near the boundary.  Doing so, one finds that near the boundary the metric has the asymptotic expansion
\begin{equation}
\label{metricasymtotics}
g_{\mu \nu}(x,r) = g^{(0)}_{\mu \nu}(x) + \dots + g^{(4)}_{\mu \nu}(x)/r^4 + \dots,
\end{equation}
where $g^{(0)}_{\mu \nu}(x)$ and $g^{(4)}_{\mu \nu}(x)$ are two independent constants 
of integration.  Via holography, the expansion coefficient $g^{(0)}_{\mu \nu}(x)$ has the physical interpretation
of the $3+1$ dimensional metric of the geometry that the dual CFT lives in and the expansion coefficient $g^{(4)}_{\mu \nu}(x)$ is related 
to the expectation value of the dual CFT stress tensor via \cite{deHaro:2000xn}
\begin{equation}
\label{stress}
\langle T_{\mu \nu}(x) \rangle = {\rm const.} \times g^{(4)}_{\mu \nu}(x).
\end{equation}

From the perspective of the dual CFT the required boundary conditions on the metric $g_{\mu \nu}(x,r)$ are obvious.  
The CFTs described by $AdS$/CFT are not gravitating: they do not backreact on the geometry they live in.
For a \textit{given} geometry that the CFT lives in, the evolution of the expectation value of the stress tensor is governed by the Hamiltonian of the CFT.
Therefore, natural boundary conditions consist of fixing $g^{(0)}_{\mu \nu}(x)$ and letting $g^{(4)}_{\mu \nu}(x)$ be dynamically determined.
The simplest possible choice is $g^{(0)}_{\mu \nu}(x) = \eta_{\mu \nu}$ where $\eta_{\mu \nu}$ is the metric of Minkowski space.
It should be emphasized that fixing  $g^{(4)}_{\mu \nu}(x)$ can lead to unstable evolution (and even the appearance of naked singularities).  
Such unphysical behavior is easy to interpret from the perspective of the dual CFT, as one must 
judiciously choose the evolution of $g^{(4)}_{\mu \nu}(x)$ so that it is consistent with the CFT Hamiltonian.

In addition to imposing the boundary condition $\lim_{r \to \infty} g_{\mu \nu}(x,r) = $ fixed, there is another type of boundary condition
that must be imposed at $r = \infty$.  Four components of Einstein's equations are radial constraint equations.  If they are satisfied at 
one value of $r$ then the other Einstein equations imply they are satisfied at all values of $r$.  In addition to imposing $\lim_{r \to \infty} g_{\mu \nu}(x,r) = $ fixed,
one must also demand that the four radial constraint equations are satisfied at $r = \infty$.  As we discuss below in Sec.~\ref{characteristic}, this is tantamount to demanding 
that the CFT stress tensor is conserved.

Even with the correct boundary conditions imposed, finding stable numerical relativity algorithms in asymptotically $AdS$ spacetime 
is a challenge.  The challenge lies in the fact that Einstein's equations are singular at $r = \infty$.  At $r = \infty$ one must not only impose boundary conditions, 
but one must also solve Einstein's equations \textit{very} well near $r = \infty$ if one is to peel off the expansion coefficient $g^{(4)}_{\mu \nu}(x)$ and obtain the expectation value of the (conserved) stress
in the dual CFT.  The simplest possible approach is to put a cutoff on the geometry at some $r = r_{\rm max}$ and impose the boundary condition
$g_{\mu \nu}(x,r_{\rm max}) = \eta_{\mu \nu}$.  However, such an approach generically does not lead to stable evolution.  
Einstein's equations imply that 
gravitational radiation propagating towards $r = \infty$ cannot change the boundary geometry at $r = \infty$
\footnote{In other words, the expectation value of the CFT stress tensor does not change the geometry that the CFT lives in.}. 
However, gravitational radiation propagating up from the bulk can change the geometry at $r_{\rm max}$.  An arbitrary boundary condition at $r_{\rm max}$ 
is generally inconsistent with Einstein's equations and thus the evolution of the stress tensor in the dual CFT.  For suitably large $r_{\rm max}$ one would
expect this effect to be small and controllable by taking $r_{\rm max}$ larger.  However, because Einstein's equations are singular at $r = \infty$, any discontinuity in the metric 
at $r_{\rm max}$ can be amplified by the singular point in Einstein's equation and lead to numerical evolution which quickly breaks down.
Therefore, one of the challenges of doing numerical relativity in asymptotically $AdS$ spacetime lies in imposing boundary conditions at $r = \infty$.  One must find an algorithm which is consistent 
with Einstein's equations and which is suitably accurate near $r =\infty$ so that the asymptotics of the metric can be peeled off and the dual CFT stress tensor can be determined.

\subsection{The characteristic formulation of Einstein's equations}
\label{characteristic}

One successful approach to numerical relativity in asymptotically $AdS$ has been 
the characteristic formulation \cite{Chesler:2008hg, Chesler:2009cy, Chesler:2010bi}. Using infalling Eddington-Finkelstein (EF) coordinates, the metric 
of asymptotically $AdS_5$ may be written
\begin{equation}
\label{efcoords}
ds^2 = -A dv^2 + \Sigma^2 g_{ij}d  x^i d  x^j + 2 F_i d x^i dv + 2 dr dv,
\end{equation}
where $v$ is EF time, $x^i$ are the CFT spatial directions and $\det g_{ij} = 1$.   Lines of constant
$v$ represent infalling null geodesics from $r = \infty$.  Moreover, the radial coordinate
$r$ is an affine parameter for these geodesics.  The metric (\ref{efcoords}) is invariant under the residual 
diffeomorphism $r \to r + \xi(v,\mathbf x)$ where $\xi(v,\mathbf x)$ is an arbitrary function.  

Einstein's equations read
\begin{equation}
R^{MN} - \frac{1}{2} G^{MN} \left ( R - 2 \Lambda \right ) = 0,
\end{equation}
where $G_{MN}$ is the $5d$ metric and $\Lambda = -6/L^2$ is the cosmological constant,
and $L$ is the $AdS$ curvature radius (which can be set to 1).
Upon substituting the metric (\ref{efcoords}) into Einstein's equations one reaches the following 
conclusions
\begin{description}
\item[1.] Time derivatives only appear in the combination $d_{+} \equiv \partial_v + \frac{1}{2} A \partial_r$.
This is the directional derivative along an outgoing null radial geodesic.
\item[2.] Given $g_{ij}$ on a slice of constant $v$, the $(0,0)$, $(0,i)$, $(0,5)$ and $(i,j)$ components of Einstein's equations 
are a system of linear ODEs in $r$ for the fields $\Sigma$, $F_i$, $d_{+} g_{ij}$ and $A$ at time $v$.  These ODEs can be integrated in from the boundary
and thereby have all boundary conditions imposed at $r = \infty$.  Once these
ODEs are solved, one can compute $\partial_v g_{ij} = d_+ g_{ij} - \frac{1}{2} A \partial_r g_{ij}$ and compute
$g_{ij}$ on the next time slice.
\item[3.]  The $(i,5)$ and $(5,5)$ components of Einstein's equations are radial constraint equations.  If they are satisfied at 
one value of $r$ the other components of Einstein's equations imply they are satisfied at all values of $r$.  These
equations must be implemented as radial boundary conditions at $r = \infty$ in the aforementioned nested linear system of ODEs.
\end{description}

To determine the requisite boundary conditions for the nested system of ODEs and how the radial constraint equations 
must be implemented it is again useful to solve Einstein's equations with a series expansion near $r = \infty$.
Doing so, one finds that the required boundary conditions on the nested system of linear ODEs consists of the 
boundary geometry \textit{and} the expectation value of the conserved densities $\langle T^{0 \mu} \rangle$ at time 
$v$.  Moreover, then one finds that the radial constraint equations are equivalent to the condition that the 
CFT stress is conserved $\partial_{\mu} \langle T^{\mu \nu} \rangle = 0$.  Therefore, the algorithm for solving
Einstein's equation in the characteristic formulation consists of 
\begin{description}
\item[1.]  Specify $g_{ij}(x,r)$ and $\langle T^{0\mu}(x) \rangle$ at time $v$.
\item[2.]  Solve the $(0,0)$, $(0,i)$, $(0,5)$ and $(i,j)$ components of Einstein's equations for 
$\Sigma$, $F_i$, $d_{+} g_{ij}$ and $A$ at time $v$ with the boundary condition that the CFT momentum density
is $\langle T^{0\mu}(x) \rangle$ and with a given boundary geometry.
\item[3.] Compute the field velocity $\partial_v g_{ij}$.  
\item[4.]  Compute the field velocity $\partial_v \langle T^{0\mu}(x) \rangle$.  This is given by the energy-momentum conservation
equation $\partial_v \langle T^{0\mu}(x) \rangle = -\nabla_i \langle T^{i\mu}(x) \rangle$, which is equivalent to the radial constraint equations.
At time $v$ the spatial components $\langle T^{ij}(x) \rangle$ are determined by the near-boundary asymptotic of $g_{ij}$.
\item[5.]  Compute $g_{ij}$ and $\langle T^{0\mu}(x) \rangle$ on the next time step and repeat the above process.
\end{description}

The utility of the characteristic formulation of Einstein's equations lies in the fact that 
Einstein's equations are integrated in from the boundary along infalling null radial geodesics.
Therefore, any numerical error made at the boundary \textit{instantaneously} falls to finite $r$ away from the singular point in Einstein's equations.
While this tames the singular point in Einstein's equations at $r = \infty$, it does not completely ameliorate it.  
One must still solve Einstein's equations very well near $r= \infty$.  Two successful approaches thus far are to i) 
solve Einstein's equations semi-analytically for $r > r_{\rm max}$ and match the semi-analytic solution onto the numerical solution at $r = r_{\rm max}$ \cite{Chesler:2008hg}
or to ii) discretize Einstein's equations using pseudospectral methods \cite{Chesler:2010bi}.  In the latter approach one can directly impose boundary conditions
at $r = \infty$, as the exponential convergence of pseudospectral methods outpaces the power law singularities in Einstein's equations.

\begin{figure}
\includegraphics[scale=0.4]{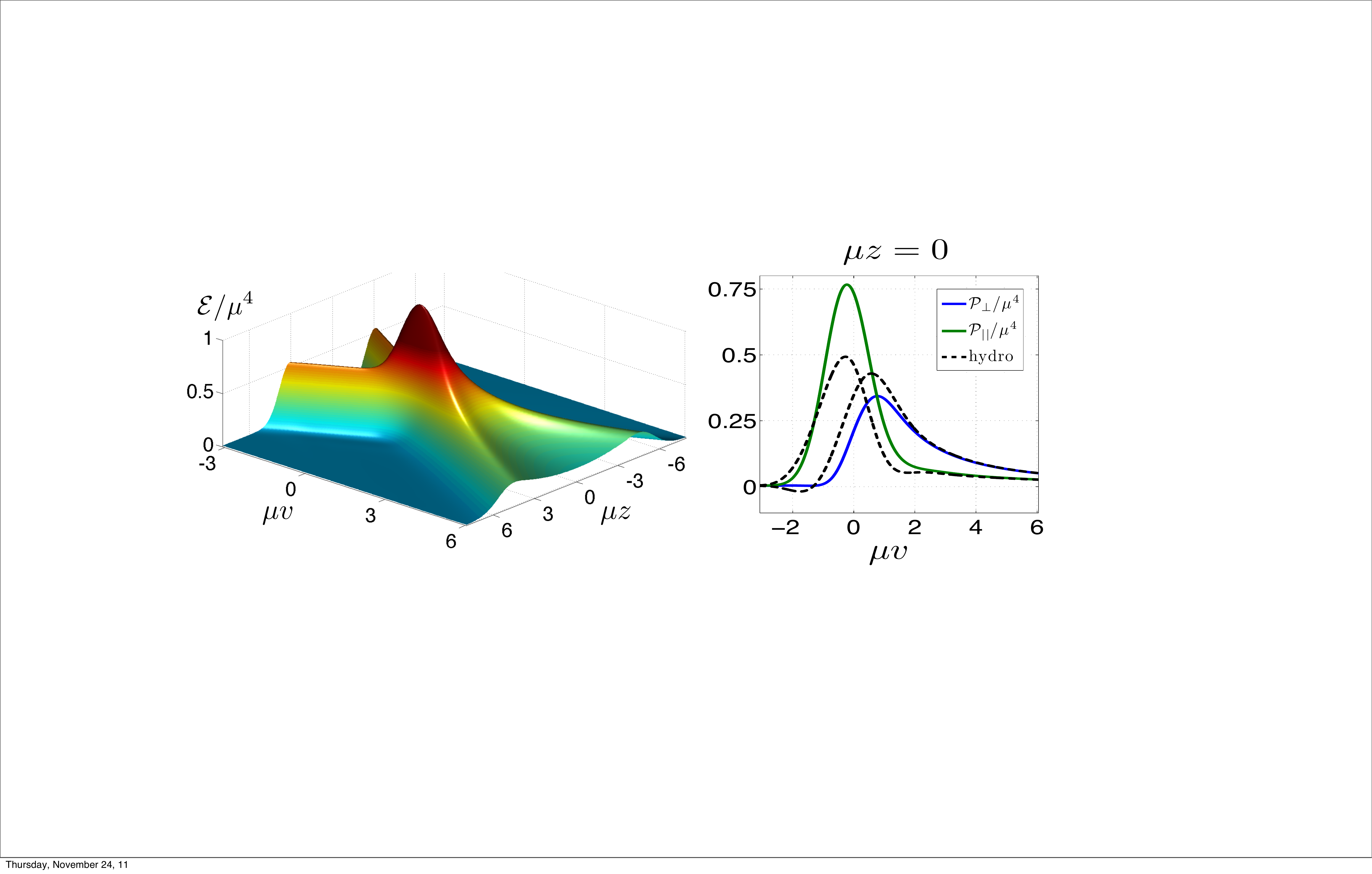}%
\caption{Left: The energy density $\mathcal E$ of two colliding sheets of matter in a holographic CFT.  Right: the transverse and longitudinal pressures $\mathcal P_{\perp}$
and $\mathcal P_{||}$ at $z = 0$.  The colliding sheets are translationally invariant
in the two directions orthogonal to the collision axis $z$.  The scale $\mu$ sets the energy density per unit area of the sheets.
The sheets propagate at the speed of light and collide at time $v = 0$.  Near $v = 0$ the system is very far-from-equilibrium and the transverse and longitudinal pressures are very different.  
However, after a few units of $1/\mu$ the dynamics of the debris left over from the collision is governed by viscous hydrodynamics with increasing accuracy as time progresses.
 \label{fig:eden}}
\end{figure}

As a proof by example that the characteristic method can be stable in asymptotically $AdS$, Fig.~\ref{fig:eden}, taken from \cite{Chesler:2010bi}, 
shows the energy density
and transverse and longitudinal pressures 
for a collision of two translationally invariant sheets of matter in a holographic CFT.  Translational invariance in the two directions 
transverse to the collision axis implies that the corresponding numerical relativity problem can be dimensionally reduced to $2+1$ dimensions.
In the dual gravitational description this problem is equivalent to the formation of a black brane via the collision of two gravitational waves.%
\footnote
  {
  We note that $AdS_5$ contains a horizon at $r = 0$, the Poincare horizon.  Because of this, a horizon exists for all times, even in the infinite past 
  before the collision event.  As a consequence of this and as a consequence of the fact that the horizon has the topology of a plane, the residual diffeomorphism invariance
  of the metric (\ref{efcoords}) can be removed by setting the position of the apparent horizon to be at $r = 1$.
  }

The fluid gravity correspondence \cite{Bhattacharyya:2008jc} implies that at suitably late times the evolution of the CFT stress tensor will be governed by hydrodynamics.  The charm of the gravitational calculation is that Einstein's equations encode 
\textit{all} physics: from far-from-equilibrium dynamics during the initial stages of the collision to the onset of hydrodynamics
at late times.  As is evident from the right panel in Fig~\ref{fig:eden}, the total time required for the collision to take place and for the system to relax to a hydrodynamic description is $\Delta_{\rm hydro} v \sim 4/\mu$, where
the scale $\mu$ is related to the energy density per unit area of the shocks $\mu^3 (N_{\rm c}^2/2 \pi^2)$ with $N_{\rm c}$ the number of colors of the CFT gauge group.
Crudely modeling heavy ion collisions at RHIC with translationally invariant sheets of matter, one can estimate $\mu = 2.3$ GeV, which yields a relaxation time $\Delta v_{\rm hydro} \sim$ 0.35 fm/c.

While demonstrating that numerical relativity in asymptotically $AdS$ geometries can be stable, the above example also demonstrates 
that thermalization times $< $ 1 fm/c are not unnatural in strongly coupled QFTs.   

\subsection{Conclusions}
\label{openissues}

The marriage of numerical relativity and holography is still in its infancy.  There are many unexplored QFT and gravity problems to study,
such as turbulence, QGP formation and thermalization in non-conformal QFTs, and Choptuik critical phenomena and its holographic interpretation.

Much work remains to be done in order to develop a robust numerical relativity toolkit applicable to holography.  While the characteristic method
discussed in Sec.~\ref{characteristic} has proven to be stable, it is not without its limitations.  One limitation is that 
the characteristic method breaks down when caustics form.  This limits the applicability of the characteristic method to black brane geometries,
where the horizon of the black brane has the topology of a plane.  Clearly, it would be desirable to develop alternative numerical relativity algorithms which lend themselves to more general geometries.

It cannot be over emphasized how valuable a robust numerical relativity toolkit would be to the study of dynamics in QFT.  The first QFTs studied were written down
a generation ago.  And yet in that generation the vast majority of the progress made in understanding dynamics in QFT has come from weak coupling expansions.
With the discovery of gauge/gravity duality, dynamics in strongly coupled QFTs is now accessible in a systematic and controlled setting.  The strongly coupled regime of QFT 
should be explored to the fullest extent possible.  While numerical relativity in holographic spacetimes may be challenging, it is vastly easier than first principles calculations in strongly coupled QFTs.

%% file: alternatives.tex
\section{Alternative theories of gravity} \label{sec:alternative}
\begin{center}
Coordinator: Leonardo Gualtieri
\end{center}

The theory of General Relativity (GR) is one of the greatest achievements in
Physics.  Almost one century after its formulation, it still stands as an
impressive construction which captures in a compelling and successful way most
of what is known about the universe on a large-scale.  And yet, both theory and
observations hint at the incompleteness of GR.

On the theoretical side, GR is conceptually disjoint from another remarkable and
successful explanation of Nature: Quantum Field Theory, which explains with
great accuracy small-scale physics.  It is believed that GR and Quantum Field
Theory should be limits of a more fundamental theory, which despite a
decades-long effort devoted to its understanding, remains as yet elusive.
Adding to this, GR is plagued by singularities and other ``weird'' phenomena
such as causality violations which appear to reveal a breakdown of the theory at
very small length scales, even though such occurrences are probably always
hidden by horizons.

On the experimental/observational side, important open problems may be
associated to deviations from GR. Among them, the most compelling regard
cosmology: observations have shown that our universe is filled with dark matter
and dark energy, which are difficult to incorporate elegantly in GR (even though
some progress has been done in this direction, see e.g.
\cite{Bernal:2009zy,Bernal:2010zz}).

One possible way out of this conundrum is to accept that Einstein's theory is
valid only in the weak field-regime, but has to be modified for a description of
strong fields.  In fact, all experiments and observations so far have only
probed the weak field regime of gravity
\cite{Bertotti:2003rm,Taylor:1982zz,Weisberg:2010zz,Kramer:2006nb} 
\footnote{Even tests from binary pulsars \cite{Taylor:1982zz,Kramer:2006nb} verify the weak
field limit of GR, because the strong gravitational field near the stellar
surface does not significantly affect the orbital motion, and the gravitational
compactness is smaller than 1 part in $10^5$ for the most relativistic (double)
binary pulsar.} and while Einstein's theory passes all these tests with flying
colours, extrapolating GR to the elusive strong-field regime may be dangerous.

Testing the strong field limit of GR is one of the main objectives of many
future astrophysical missions \cite{nasa,esa}. In the context of testing
alternative theories of gravity, arguably the most promising of these missions
is the space-based gravitational wave (GW) detector LISA, which is expected
(even in E.U.-redefined configuration \cite{AmaroSeoane:2012km}) not only to
directly observe gravitational waves, but to give birth to the so-called
``gravitational-wave astronomy'', i.e., observing the universe through a new
window, the most appropriate to look at violent, strong field phenomena
occurring in our universe.  Several other projects to probe the strong field
regime of GR are currently in construction or under design study. These include:
the second generation ground-based GW detectors Advanced LIGO/Virgo
\cite{ligovirgo}, and, on a longer timescale, LCGT and AIGO \cite{lcgtaigo}; the
third generation detector ET \cite{et}; more advanced space-based detectors
(DECIGO, BBO) \cite{Kawamura:2011zz,Crowder:2005nr}; next generation X-ray and
microwave detectors to study black holes and the cosmic microwave background
(IXO, NICER) \cite{xray}. Further experiments target the observation of specific
violations of GR, like violations of Lorentz symmetry, of the equivalence
principle, of the expected polarization and speed of GWs, etc.  We expect then
to have soon a large amount of data, which will allow us to test, for the first
time, the strong field limit of gravity.  A signature of new physics in these
experiments and observations could be a ``message in a bottle'' coming from a
more fundamental theory, standing at energies far beyond our reach.

In this context, it is not surprising that the community of theorists has been
paying, in recent years, more and more attention to possible deviations of GR,
and more generally to alternative theories of gravity. Many theories have been
proposed, such as scalar-tensor theories \cite{Fujii2003}, scalar-tensor-vector
theories \cite{Bekenstein:2004ne}, massive graviton theories
\cite{Goldhaber:1974wg,Dubovsky:2004sg}, brane-world models
\cite{Randall:1999ee,Randall:1999vf,Dvali:2000hr}, f(R) theories
\cite{Sotiriou:2008rp}, quadratic curvature corrections (e.g. Gauss Bonnet
gravity \cite{Zwiebach:1985uq}, Chern-Simons gravity \cite{Alexander:2009tp})
or Lifshitz-type theories \cite{Horava:2009uw}.
Given this plethora of possible corrections to GR, it is crucial to devise some guideline
which allows us to make contact between theoretically conceivable models and
upcoming observations.

In the following we discuss some important issues about alternative theories of
gravity in four dimensions, with the aim to report both on the
state-of-the-art and on the ongoing discussion on these topics. We will not
discuss here alternative theories of gravity in $D>4$; we only mention that some
effort has been done to study deviations from GR in higher dimensions, mainly
in the context of theories with quadratic curvature corrections
\cite{Lovelock:1971yv,Zegers:2005vx,Charmousis:2008kc}.

\subsection{Which kind of GR deviations should we expect?}

An answer, even partial, to this question would be extremely valuable.  Indeed,
on the one hand, the realm of possible alternative theories of gravity is too
large to be the starting point for systematic tests of GR. On the other hand,
choosing one specific alternative theory of gravity could become a merely formal
exercise, unless the choice is made on physical grounds.  Actually, many
alternative theories of gravity claim to be ``string inspired'', since they have
features which also appear, in different contexts, in String Theory/M Theory
(ST). This is the case, for instance, for scalar-tensor theory \cite{Fujii2003},
in which, as in ST, the spacetime metric is non-minimally coupled to a scalar
field; of theories with quadratic curvature corrections
\cite{Zwiebach:1985uq,Alexander:2009tp}, which arise in several low-energy
truncations of ST; of brane-world theories where, as in ST, gauge fields can be
confined on submanifolds (branes) of the spacetime; and so on.

These are merely inspirations and analogies, however. It would be extremely
useful, in the study of possible GR modifications, to have a guidance from
candidate fundamental theories (ST, loop quantum gravity, etc.), but actual ST
constraints on the low energy theory are still poor, because we do not know the
symmetry breaking mechanism which would yield our world. Many deviations from GR
can be accommodated in some model of ST. Actually, it would be more appropriate
to speak of ``predictions of a ST model'' rather than ``ST predictions'', since
today ST should be viewed as a framework on which different models can be
defined, rather than a well defined theory.

Given that ST is one of the few candidates for a unified theory, it is perhaps
wise to extract from it general indications, which may be useful ({\it if} it is
the correct route to the fundamental theory unifying GR and quantum field
theory) as a possible, preliminary guideline.  In this context, the most likely
and potentially interesting GR corrections would be associated to couplings with
other fields (scalars, vectors, etc.)  Another interesting possibility is that
GR corrections would involve curvature invariants. The dimensionful coupling
constant of such curvature invariants may, however, be suppressed by the Planck
scale, making these corrections irrelevant for astrophysical phenomena. On the
other hand, both the hierarchy problem and the cosmological constant problem
suggest that possible couplings to gravity may be very different from those
expected from standard field theory. Thus, if for some reason these corrections
are not Planck-suppressed, alternative theories like Chern-Simons gravity or
Gauss-Bonnet gravity may have important implications for gravitational-wave
astronomy.

Alternatively, one could adopt an ``agnostic'' attitude, not trying to infer
general indications from candidate fundamental theories, and instead considering
whether gravitational wave observations or other gravitational experiments can
provide any constraints on gravity theories beyond that of Einstein.  One route
toward this goal is to think about what possible symmetries or fundamental GR
principles one could compromise to study whether nature violates them. This is
the case, for instance, of generic breaking of parity invariance (captured in
Chern-Simons gravity \cite{Alexander:2009tp}), or of kinematical violations of
Lorentz-Symmetry in the propagation of gravitational modes
\cite{Mirshekari:2011yq}.  Whether one expects such deviations from fundamental
principles or not, it might still be worthwhile to consider what current or
future experiments have to say about these topics.

\subsection{Strategies to find a GR deviation in experiments and observations}

The answer to this question is not unique. A possible strategy may be to build a
parametrization as general as possible of metric theories of gravity, modelling
in this framework the relevant (strong field) astrophysical processes, and
determining observable signatures in terms of the parameters of the theory.
This could be made in two ways (which have been dubbed, respectively, top-down
and bottom-up \cite{Psaltis:2009xf}): parameterizing the action and looking at
the phenomenological consequences, or parameterizing a phenomenological
description of the observations, and inferring the consequences on the
underlying theory.  An attempt to pursue the former approach has been made in
Ref.~\cite{Yunes:2011we}, where a parametrization of the action in terms of
polynomials in the curvature tensor has been proposed. An example of the latter,
instead, is the parametrization of the black hole spacetime (in a general theory
of gravity) in terms of multipole momenta, which has been proposed in the
context of extreme mass-ratio inspirals \cite{Ryan:1995wh,Vigeland:2011ji}.
Another example is given by the ppE formalism
\cite{Yunes:2009ke,Cornish:2011ys}, in which a parametrization of the the
  gravitational waveform emitted in black hole binary coalescences is proposed.
All these approaches are attempts to build a general parametrization of the
gravitational theory, to be compared with observations.

On the other hand, a general parametrization of gravitational theories could be
difficult to implement in the analysis of data (for instance from gravitational
interferometers). Indeed, it could depend on too many parameters to preserve
practicality of the matched filtering process: too many free parameters in the
template space can introduce degeneracies in the parameter estimation process
and also lead to larger uncertainties in those estimations, i.e., they
  could raise the false-alarm rate. It should further be remarked that having a
precise template will be crucial for the determination of the physical
parameters, but should be less important for detection, \cite{Cutler:1992tc}.
Therefore, it will probably be the outcome of the experiments which
dictate how we should proceed or which strategies will work best in testing the
behaviour of the gravitational interaction. 

Nevertheless, there exists strong motivation for embedding the alternative
theories of gravity proposed so far in a large class of theories. A clever
strategy could be to first find the general signature of a theory (or of a class
of theories), then identify the experimental/observational setup in which such a
signature is enhanced. Thus, the search for ``smoking guns'' of alternative
theories (for instance spontaneous scalarization \cite{Damour:1992we} and
floating orbits \cite{Cardoso:2011xi} in scalar-tensor theories, birefringence
of GWs in parity-violating theories \cite{Lue:1998mq}) would be a solid route to
establish or rule out many candidate theories. Knowing ``where to look'' would
also be extremely useful to conceive new experiments and to fine-tune the
experiments now in the commissioning and building stages. We would be able, once
a deviation from the GR prediction is detected or observed, to understand
its theoretical implications, and translate the new observation into a deeper
understanding of the fundamental laws of nature.

It is also worth saying that, even in the absence of any observational evidence
to date for deviations from GR, experimental/observational data are already
providing useful information on alternative theories of gravity. Indeed,
negative results (even in weak field processes) enable us to constrain the set
of allowed theories
\cite{Talmadge:1988qz,Bertotti:2003rm,Will2006,Yunes:2009hc,Everitt:2011hp,Arvanitaki:2010sy}.

\subsection{GR deviations in astrophysical processes}

To test deviations from GR we should look at the most violent astrophysical
processes dominated by gravitational fields, which are the coalescence of
compact binary systems formed by black holes and/or neutron stars, either of
comparable masses \cite{Will:1997bb,Stavridis:2009mb,Berti:2004bd} or of extreme
mass ratio \cite{Sopuerta:2009iy,Sopuerta:2010zy,Yunes:2011aa}. Black hole
oscillations are a promising process, too, since they encode in a clean way the
features of the underlying theory of gravity \cite{Berti:2005ys}. Such processes
provide an optimal testbed for GR, as they probe the pure gravitational
interaction. Indeed, most of the literature on phenomenological bounds of
alternative theories of gravity refers to such processes.

Possible insights from black hole and neutron star physics would be
complementary to the large amount of literature on alternative theories in
cosmological scenarios. Indeed, cosmological observations could also provide a
valuable tool to study alternative theories of gravity. For instance, some of
these theories (braneworld models, $f(R)$ gravity) could account for the
acceleration of the universe without the need of including dark energy
\cite{Koyama:2005kd,Kunz:2006ca,Carroll:2006jn}; these models can then be tested
against the large amount of observational data on supernovae, cosmic microwave
background, gravitational lensing, galaxy clustering, etc. \cite{Wang:2007fsa}
(see also \cite{Koyama:2009me}). Other theories (such as scalar-vector-tensor
theories) could explain the galaxy rotation curves without the need of including
dark matter \cite{Bekenstein:2004ne} (see also \cite{Skordis:2008pq}). This
topic is beyond the scope of this article; for a comprehensive review we refer
the reader to \cite{Clifton:2011jh}.

Some of these processes, however, especially those involving black holes, do not
probe a key aspect of GR, namely the coupling of gravitation to matter. Possible
constraints in this sector may come from different kinds of strong-gravity
processes, like those involving the internal dynamics of neutron stars.
Unfortunately, the uncertainty about the behavior of matter at extreme density
in the core of a neutron star makes it difficult to disentangle the effects of
an alternative theory of gravity from those due to a different equation of
state. Within the next few years, astrophysical and gravitational wave
observations may be able to shed more light onto the neutron star equation of
state.  This will open up the possibility of making precision tests of the
coupling of gravity with matter \cite{Sotani:2004rq,Sotani:2005qx,Pani:2011mg}.
Accretion of matter into black holes has also been studied in alternative
theories of gravity \cite{Campanelli:1993sm,Bambi:2011vc}, but the
uncertainties in our knowledge of this process make it difficult to use it as
a testbed of GR against other theories.

More generally, rather than constraining small corrections to standard gravity,
it could be useful to look for effects which only appear in alternative
theories, but identically vanish in GR
\cite{Damour:1992we,Cardoso:2011xi,Lue:1998mq}. Such effects may provide clear
observational signatures and thus prove effective in constraining or ruling out
alternative theories.

\subsection{Numerical relativity and alternative theories of gravity}

Numerical relativity may be a powerful tool to study alternative theories of
gravity, and to understand their phenomenology in strong field astrophysical
processes. Numerical simulations, however, have so far almost exclusively been
restricted to highly symmetric configurations such as the spherically symmetric
gravitational collapse of dust \cite{Novak1998, Novak1998a, Novak2000}, or
processes involving scalar fields in spherical symmetry
\cite{Contaldi:2008iw}. We note, however, the recent study of black holes in
scalar-tensor theory by Healy {\em et al.} \cite{Healy:2011ef}. With that
exception, though, the use of numerical relativity following the 2005
breakthroughs \cite{Pretorius:2005gq,Baker:2005vv,Campanelli:2005dd} in the
study of alternative theories of gravity remains essentially uncharted
territory. Aside from the rather well-studied case of scalar-tensor theories
(see also \cite{Salgado2006,Salgado2008}), it is not even clear to what extent
the stability properties of the present state-of-the-art codes in $3+1$ and
higher-dimensional spacetimes carry over to modifications of general
relativity. Furthermore, for those cases where the action contains higher-order
polynomials of the Riemann tensor, the existence of a well-posed initial value
formulation of the field equations is not known. For this class of theories,
which includes for example Chern-Simons Gravity, the field equations contain
derivatives higher than second order which may drastically change the
mathematical structure of the theory (see e.g. \cite{Motohashi:2011ds}).

There may be reason, however, for some optimism in this regard. We emphasize
that the following items cannot be regarded as mathematical arguments, but
merely as an intuitive motivation for carrying out such investigations with some
level of confidence. First, if a given theory correctly describes real physical
processes, we should expect there to exist a well-posed form of this theory.
Second, we typically consider scenarios which only mildly deviate from GR. Given
that GR itself has well-posed initial-value formulations, we might hope that
this remains the case, at least in a ``neighborhood around GR in the space of
theories''.  Third, modified theories of gravity often arise as the low-energy
limits of more fundamental theories. It could then be acceptable
if a modified theory is well-posed in certain regimes only.

Deriving fully non-linear numerical evolutions in the framework of alternative
theories of gravity will require substantial effort, but the wealth of
physical systems thus opened up for systematic study almost certainly
justifies the effort. For instance, we would be able to model
processes with low degree of symmetry or those involving complex
matter distributions and/or other fields, whose study is not possible
using semi-analytic approaches such as perturbation theory or parametrized
post-Newtonian expansions.

The answer to the above questions and issues crucially depends on the specific alternative
theory to be studied. Two recent works on different directions allow for optimism.
In the case of GR coupled with a scalar and/or vector field, the mathematical structure of the evolution equations is expected to be
preserved \cite{Salgado2006,Salgado2008}. In line with this expectation, numerical relativity
simulations of binary black hole coalescences in scalar/tensor theory have
recently been performed \cite{Healy:2011ef}.
A second exciting development concerns a well-known conjecture \cite{Tanaka:2002rb,Emparan:2002px,Emparan:2002jp} that large static black holes do not exist in type II Randall-Sundrum scenarios of modified gravity model. These works used indirect arguments to claim static solutions do not exist, and gravitational collapse would yield a dynamical black hole that would ``evaporate'' classically due to (classical) gravitational radiation. A counter-example to this conjecture was recently presented where static black holes were numerically generated \cite{Figueras:2011gd,Figueras:2011va}.

%% file: Approximation.tex
\section{Approximation Methods in GR}\label{sec:approximation}
\begin{center}
Coordinator: Carlos F. Sopuerta
\end{center}

\subsection{Relativistic Perturbation Theory}
In the theory of General Relativity (GR) gravity does no longer appear as a 
force but as a manifestation of the geometry of spacetime, which is a dynamical entity.  
As such, the gravitational
dynamics is encoded by the spacetime metric tensor, a spin-2 field that 
satisfies the Einstein field equations.  The main difficulties that arise in GR
are related to the  non-linearity  and the diffeomorphism invariance of the theory,
which complicate solving the field equations and understanding the solutions.
Most of what we have learned about GR comes from the detailed studies of a few exact solutions (see, e.g.~\cite{Hawking:1973uf,Stephani:2003tm}), 
in particular Minkowski, Schwarzschild, Kerr, Friedmann-Lema$\hat{{\rm i}}$tre-Robertson-Walker (FLRW), etc.).
These exact solutions can also be taken as the basis for the study of more 
complex physical situations, e.g. oscillations of compact stars and Black Holes (BHs),
cosmological structure formation, etc.  The reason for this is that many of these physical
situations admit a perturbative analysis where the zeroth-order solution (usually
called the background geometry) is one of the exact solutions mentioned above.  
The starting point of relativistic perturbation theory is to consider two different
spacetimes, the \emph{physical} one, which describes the actual physical system, and
the \emph{background} one, which corresponds to a simpler idealized situation.  
We can relate these two spacetimes using different maps (which identify points of the background
and physical spacetimes and can be used to transport the tensorial structure between them), 
each of them corresponding to a \emph{gauge} choice and the transformation between maps is 
known as a gauge transformation. Using one such mapping we can write the metric tensor of 
the physical spacetime, $\met^{}_{\mu\nu}$, as
\begin{equation}
\met^{}_{\mu\nu} = \bmet^{}_{\mu\nu} + \delta\met^{}_{\mu\nu}  \qquad
(\mu,\nu = 0\,,\ldots\,,D-1)\,,
\end{equation}
where $D$ is the spacetime dimension, $\bmet^{}_{\mu\nu}$ denotes
the background spacetime metric and $\delta\met^{}_{\mu\nu}$ are the
metric perturbations, which can be split into a first-order piece (that
satisfies the Einstein equations linearized with respect to the background
metric), a second-order piece, etc.  Experience tells us that the
gauge freedom in the identification between the background and physical
spacetimes can cause problems, in fact many authors talk about the {\em
gauge problem}; e.~g.~\cite{Sago:2002fe,Barack:2005nr}.
These problems arise in the physical interpretation
of results, and in particular when one works with a family of gauges
instead of a unique gauge. For instance, considering only first-order
perturbations (something similar applies to higher perturbative orders),
the Lorenz gauge condition $\tilde\nabla^{\mu}(\delta\met^{}_{\mu\nu} -
(1/2)\bmet^{}_{\mu\nu}\bmet^{\rho\sigma}\delta\met^{}_{\rho\sigma})
= 0$, does not identify a unique gauge since we
can perform gauge transformations whose generating
vector field $\xi^{\mu}$ satisfies $\Box \xi^{\mu}\equiv
\tilde{\met}^{\rho\sigma}\tilde{\nabla}^{}_{\rho}\tilde{\nabla}^{}_{\sigma}\xi^{\mu}=0$.
These gauge transformations respect the Lorenz gauge condition
but can change the metric perturbations and hence one must be
careful when dealing with perturbations in this family of gauges.
To avoid these potential problems, a usual procedure is to look
for gauge-invariant quantities, i.e. quantities that have the same
values independently of the gauge one is working in.  For details
on the mathematical formulation of relativistic perturbation theory
see~\cite{1974RSPSA.341...49S,Bruni:1996im,Bruni:1999et,Bruni:2002sm,Sopuerta:2003rg,Brizuela:2006ne,Brizuela:2010qu}.
In the case of $4$D spacetimes, it has been applied and developed for
different physical systems, mainly for the following: \\
(i) Perturbations of flat spacetime (see, e.g.~\cite{Misner:1973cw}).
These are the most simple type of perturbations that one can think of
since the background spacetime geometry is the Minkowski flat spacetime.
They are useful for several types of studies, in particular to describe
the propagation of gravitational waves far away from the sources. Perturbations of flat spacetime is also at the core
of the so-called post-Minkowskian approximation,
where one is interested in gravitational weak field phenomena.\\
(ii) Cosmological Perturbation
Theory~\cite{Lifshitz:1946el,bardeen:1980cp,Brandenberger:1983rb,Kodama:1984hk,Ellis:1989eb,Mukhanov1992203,Bruni:1993mb,Malik:2009ka}.
As the name indicates, here one perturbs a spacetime of cosmological
character.  Most of the work done in the literature has focused on
the development of perturbations of the FLRW cosmological models.
They are one of the cornerstones of the present standard model
of cosmology as they are fundamental for the description of many
physical phenomena in cosmology: The formation and growth of
structures in the early universe within the inflationary paradigm,
the description of anisotropies in the cosmic microwave background
(e.g. the Sachs-Wolfe effect~\cite{Sachs:1967er}), etc.  Recently,
second-order perturbation theory has been used to argue that the
acceleration of the Universe (the so-called dark energy problem)
could be explained as a backreaction effect (for a review see,
e.g.~\cite{Buchert:2007ik}), although some other studies indicate that
this is unlikely~\cite{Ishibashi:2005sj,Kumar:2008uk}.\\
(iii) BH Perturbation Theory.  Here we have to distinguish
between non-rotating (Schwarzschild) and rotating (Kerr)
black holes (for textbook reviews on BHs and BH perturbations
see~\cite{Chandrasekhar:1992bo,Frolov:1997fn}).  In the case of
non-rotating black holes the formalism has been developed based on metric
perturbation variables~\cite{Regge:1957rw,Zerilli:1970fj,Moncrief:1974vm}.
Spherical symmetry is a key ingredient since it allows decomposition
in tensor spherical harmonics and the equations for each harmonic
can be decoupled in terms of gauge-invariant master functions
that satisfy wave-type equations in 1+1 dimensions.  Once these
equations are solved, all metric perturbations can be reconstructed
from the master functions.  The situation in the rotating case
is significantly different.  First of all, partly due to the lack
of the spherical symmetry, we do not have any metric-perturbation
based formalism to decouple the first-order perturbative equations.
Nevertheless, we do have a formalism based on curvature variables
due to Teukolsky~\cite{Teukolsky:1972le,Teukolsky:1973ap} that
provides master equations (the Teukolsky equation) for the Weyl
tensor components that can be associated with ingoing and outgoing
gravitational radiation.  In both cases, rotating and non-rotating,
we can compute gravitational waveforms, and energy and momentum
fluxes radiated at infinity from the master functions.  What remains
to be done in the rotating case is to establish, in general, the
reconstruction of the metric perturbations from the curvature-based
variables, although significant progress has
already been made
(see~\cite{Chrzanowski:1975wv,Wald:1978vm,Kegeles:1979an,Lousto:2002em}).

BH perturbation theory has been applied to study the stability of
BHs~\cite{Vishveshwara:1970vi,1987CQGra...4..893K,Whiting:1988vc},
the computation of quasi-normal modes of
BHs~\cite{Vishveshwara:1970na,Press:1971qn,Chandrasekhar:1975qn},  etc.
In an analogous way to the BH case, perturbations and quasi-normal
oscillations of relativistic stars have been extensively studied
(see~\cite{lrr-1999-2} for a review).  It also has been applied to the
description of the dynamics of binary systems with an extreme mass
ratio and their gravitational wave emission, the so-called {\em
Extreme-Mass-Ratio Inspirals} (EMRIs). This is a very demanding
subject in terms of the perturbation theory technology needed for
the computations.  Usually one describes the small compact object as
a point mass (which is at odds with the full theory but allows us some
simplifications) that induces perturbations on the geometry of the large
one, considered to be a (supermassive) black hole.  Then, the inspiral
can be described in terms of the action of a local force, called the
{\em self-force}, that can be constructed from the gradients of the
first-order perturbations. However, the point-like description of the
small compact object leads to singularities in the perturbative solution
that must be regularized.  Procedures to regularize the solutions have
been devised in the Lorenz gauge (see, e.g.~\cite{Barack:2001bw}) but
working in this gauge complicates the computations as we no longer
have some of the advantages associated with the well-known Regge-Wheeler
gauge in the case of non-rotating BHs, as for instance the decoupling
of the metric perturbations; something similar happens in the
case of spinning BHs.  At present, the gravitational self-force has
been computed for the case of a non-rotating BH
first using time-domain techniques~\cite{Barack:2009ey,Barack:2010tm}
and later with frequency-domain techniques~\cite{Akcay:2010dx}.  These
calculations have allowed the study of some physical consequences of the
self-force~\cite{Barack:2010ny,Barack:2011ed,LeTiec:2011bk}. Progress
is being made towards calculations for the case of a spinning
BH~\cite{Shah:2010bi}.  In any case, given the amount of cycles
required for EMRI GWs (it scales with the inverse of the mass ratio,
which can be in the range $10^{-7}-10^{-3}$) we cannot expect to generate
complete gravitational waveform template banks by means of full self-force
calculations.  Instead, the goal of these studies should be to understand
all the details of the structure of the self-force so that we can
formulate efficient and precise algorithms to create the waveforms needed
for gravitational-wave observatories like LISA~\cite{Danzmann:2003tv}
(see~\cite{Poisson:2004lr,Barack:2009ux,Thornburg:2011ei} for reviews
on the progress in the self-force program).  Observations of EMRIs have
a great potential for improving our understanding of BHs and even the
theory of gravity (see, e.g.~\cite{Sopuerta:2010zy}).

Usually approximation methods require the introduction of a smallness
parameter to establish in which sense the perturbations are small.
In relativistic perturbation theory this is done in an implicit
way, in the sense that we could replace $\delta\met^{}_{\mu\nu}$ by
$\lambda\delta\met^{}_{\mu\nu}$, where $\lambda$ is a formal perturbation
parameter without having a direct physical meaning (as in cosmology,
in backreaction problems, or in the study of quasi-normal modes of stars
and black holes), although there are situations in which we can assign
to it a specific physical meaning, as in the study of black hole mergers
via the close limit approximation, in the analysis of quasi-normal mode
excitation by a physical source, or in the modeling of perturbations
generated by the collapse of a rotating star.

From these basic ideas one can develop other 
approximation schemes in GR.  In general, this involves adding
extra assumptions in the approximations, combining different
schemes, etc.  One example of this is post-Newtonian  theory~(see,
e.g.~\cite{1987thyg.book..128D,Blanchet:2002av,poisson-lecture-notes}).
In general, a post-Newtonian approximation can describe the GR dynamics
in the regimes where the speeds involved are smaller than the speed
of light ($v \ll 1$) and the gravitational interaction is {\em weak}
($M/R\ll 1$, where $R$ denotes the size of each body
or the typical orbital separations).  Moreover, it is well-known that
the post-Newtonian approximation is valid only in  the vicinity of the
massive objects (a region around the bodies that is small as compared
with the wavelength of the gravitational waves emitted by the system).
One can use a post-Minkowskian approximation (where only the weak
gravitational field condition $M/R \ll 1$ is imposed) to describe the
gravitational field outside the near zone.  Then, one can match the two
expansions, the post-Newtonian and post-Minkowskian ones, by means of
the method of matched asymptotic expansions, which is a key ingredient
to connect the orbital motion and the gravitational-wave emission.
This process is quite involved in practice and has been developed
during many years (see~\cite{1987thyg.book..128D} for a review).
The result is a framework in which we can in particular model the
general relativistic dynamics of compact binary systems, including
their gravitational-wave emission.  By comparing with recent numerical
relativity simulations, it has been found that post-Newtonian
computations provide a good description of inspiral of a binary
system close to the merger phase.
In order to improve these results new
schemes that use post-Newtonian theory has been proposed.  In particular,
the Effective-one-Body scheme~\cite{Buonanno:1998gg,Buonanno:2000ef}
has been developed to a point where one can construct gravitational
waveforms for the whole binary BH coalescence~\cite{Damour:2009ic},
including merger and ringdown in the case of non-rotating black holes;
for first applications of the Effective-one-Body
method to spinning binaries see also Refs.~\cite{Pan:2009wj,Barausse:2009xi}.

Another relativistic computation scheme that has been useful
in General Relativity is the so-called {\em Close-Limit}
approximation~\cite{Price:1994pm}.  This is based on the realization
that binary black hole coalescence can be divided in three stages:
a long and relatively slow inspiral, a short non-linear merger phase,
and finally the ringdown of the final black hole towards a stationary
black hole state.  Then, it turns out that in last two stages of this
process (or at least including a part of the merger phase where we are
close to the formation of an apparent horizon) the binary black hole
system can be seen as a single deformed black hole.  The idea is then
to map the two-black-holes system to a single black hole (we must read
its mass and eventually its spin from this mapping) plus perturbations
(which are also read from the mapping and this is the key part of the
computation).  Then, by using the standard techniques of BH perturbation
theory that we have described above we can evolve the system until the
final BH is settled, and from this evolution we can compute physical
quantities such as energy and momentum emission in gravitational waves.
It has been shown~\cite{Anninos:1993zj,Anninos:1995vf} that in the case
of head-on collisions of BHs, the Close Limit approximation provides
accurate results in comparison with numerical relativity results.
The Close Limit approximation scheme has been developed by a number of
authors~\cite{Gleiser:1996yc,Gleiser:1998rw,Krivan:1998er,Nicasio:1998aj,Khanna:1999mh,Khanna:2000dg,Gleiser:2001in,Khanna:2001ch,Sarbach:2001tj,Khanna:2002qp}
and it has been applied to astrophysical problems like the
computation of recoil velocities
(see e.~g.~\cite{Campanelli:2007ew,Gonzalez:2007hi})
of BHs formed in binary BH
collisions, in the past for head-on collisions~\cite{Andrade:1996pc}
and recently for non-rotating BHs in circular and in small
eccentricity orbits~\cite{Sopuerta:2006wj,Sopuerta:2006et} (see
also~\cite{LeTiec:2009yf,LeTiec:2009yg}).  Since the Close Limit
approximation scheme applies to the final part of the binary BH
coalescence it offers considerable potential for problems in High-Energy
Physics that involve black hole collisions or similar systems (see
next subsection). In essence, the close limit approximation is a linear approximation around the final, equilibrium black hole state. In the context of gauge-gravity duality, it has been recently applied in \cite{Heller:2012}.

Up to now we have described approximation methods that use the
most common background spacetimes and the main perturbative
schemes that have been developed from them to study a variety of
physical phenomena.  However, this is not all what has been done.
There are other 4D spacetimes that have received attention
and perturbations of them have been studied.  Among
those, we note the
study of oscillations (quasi-normal modes) of Schwarzschild-dS
black holes~\cite{Cardoso:2003sw}, Schwarzschild-$AdS$ black
holes~\cite{Horowitz:1999jd,Cardoso:2001bb,Cardoso:2003cj,Bakas:2008gz},
Reissner-Nordstrom-$AdS$ black holes~\cite{Berti:2003ud} and Kerr-$AdS$ black
holes~\cite{Giammatteo:2005vu}. Oscillations of black holes in alternative
theories of gravity, such as higher derivative gravity, have also been
studied~\cite{Moura:2006pz,Kanti:1997br,Torii:1996yi,Pani:2009wy,Molina:2010fb}
(for a review, see \cite{Berti:2009kk}).

\subsection{Perturbations of Higher-Dimensional Spacetimes}
All that we have described until now is standard relativistic perturbation
theory in the case of $4$D spacetimes.  Nevertheless,  it is clear that
the basic ideas and foundations of relativistic perturbation theory
can be transferred without problems to higher dimensional spacetimes.
Here, we will only consider higher-dimensional spacetimes governed
by Einstein's equations, $G^{}_{\mu\nu}[g^{}_{\rho\sigma}] = 8\pi\,
T^{}_{\mu\nu}$, where $G^{}_{\mu\nu}$ is the Einstein curvature tensor
and $T^{}_{\mu\nu}$ is the stress-energy tensor.

In the last decades,
there has been growing interest in physical phenomena in
higher dimensions motivated by the emergence of
new theoretical models in High Energy Physics.  In particular on
theoretical models either based on string theory or motivated by it.
An interesting feature of many of these scenarios is that they involve
higher-dimensional spacetime geometries where the extra dimensions
need not be compactified or have a large curvature radius.  Many of
these theories deal with physical situations that involve energies
beyond those associated with the standard model, and given that some
of them suggest that the {\em fundamental} Planck mass may be small (
as low as the TeV range, which can be a solution to the so-called {\em hierarchy
problem}~\cite{ArkaniHamed:1998rs}), black holes and other {\em dark}
objects (containing horizons) in spacetimes with different number
of dimensions can play an important role and as such are an important
subject of investigation (see Sections \ref{sec:nr}, \ref{BHs} and \ref{sec:tp} for further details). 
There is also a strong motivation coming from the correspondence between $\mathcal{N} = 4$ Super Yang-Mills theory in
the large $N$ limit and type-IIB string theory in $AdS_{5}\times S^{5}$,
the $AdS$/CFT correspondence~\cite{Maldacena:1997re}.   The idea is that
in the low energy limit, string theory reduces to classical supergravity
and the $AdS$/CFT correspondence becomes a tool to calculate the gauge
field-theory correlation functions in the strong coupling limit leading
to non-trivial predictions on the behavior of gauge theory fluids,
which has a lot of applications that have made this correspondence one
of the main subjects of current research; see also Sec.~\ref{sec:holography}.

Cosmological perturbations have also been analyzed in certain theoretical
scenarios, as for instance in braneworld models.  These models involve
at least one extra dimension.  Two scenarios where these cosmological
perturbations have been studied extensively are the Randall-Sundrum
type II scenario~\cite{Randall:1999vf} (see, e.g.~\cite{Kodama:2000fa})
and a closely-related one, the Dvali-Gabadadze-Porrati
model~\cite{Dvali:2000hr}.  Again, the interest here is in the early
Universe dynamics where the energies can make these models deviate
from the usual GR cosmological dynamics (see~\cite{Maartens:2010ar}
for a review).

In higher dimensions many geometrical properties that are
valid in 4D no longer hold (and some hold in an appropriate
form~\cite{Hollands:2006rj,Hollands:2008wn}).  In the context of
High Energy Physics, it is particularly relevant that the
topology of connected components of spacelike sections of event
horizons does not need to be that of a sphere, as happens in
4D~\cite{Hawking:1971vc,Hawking:1973uf} (see also~\cite{Hollands:2010qy}).
This opens the door to many topologically different objects with horizons
\cite{Emparan:2008eg}.  Not all these objects need to be stable,
and actually many of them are not, and this strongly motivates the
study of deviations from all those geometries;
cf.~the discussion of stability in Sec.~\ref{BHs}.
Therefore, despite the fact that the
basics of relativistic perturbation theory are the same independently of
the dimensionality,
the development of a complete perturbative scheme in the
sense of constructing gauge-invariant quantities and their equations for
describing the physics associated with the perturbations is a task that
needs to be done for each background spacetime geometry or at least
for families of background spacetime geometries.

As we have discussed above, in 4D
the perturbative schemes for BHs are quite developed. For higher
dimensional black holes there is still a long way to reach a complete
formulation for perturbations.  Nevertheless, significant progress
has already been made.  For static black holes in arbitrary
higher dimensions a set of decoupled master equations, which
correspond to the Regge-Wheeler-Zerilli equations in 4D, have been
found~\cite{Kodama:2003jz,Kodama:2003kk,Ishibashi:2011hk}. In this
case, for a $D=2+d$ BH background geometry, where $d$ is the number of
dimensions of the internal space (which is an Einstein space; in $D=4$
$\Leftrightarrow$ $d=2$, it is the 2-sphere).  Then, the perturbative
variables are classified according to their tensorial behaviour
on the internal space and gauge-invariant variables are introduced.
Furthermore, for each type of perturbations, decoupled master equations
are found for scalar master functions on the 2-dimensional sector
of the BH background spacetime.  Using these techniques, it has been
established~\cite{Ishibashi:2003ap} that the Schwarzschild solution is
mode-stable against linearized gravitational perturbations for all dimensions
$D > 4$. More specifically, it was shown that the master equation
for each tensorial type of perturbations does not admit normalisable
negative-modes which would describe unstable solutions. It was also shown
that there exists no static perturbation which is regular everywhere
outside the event horizon and well-behaved at spatial infinity, which
is a check, within the perturbation framework, of the uniqueness of the
higher-dimensional spherically symmetric, static, vacuum black hole.

The situation is different for the rotating case, where we have different
families of solutions. So far, studies of linearized perturbations of
higher-dimensional rotating black holes have exploited isometries of
black hole spacetimes, e.g. enhancement of symmetry of the Myers-Perry
family of solutions~\cite{Myers:1986un} by choosing some of the intrinsic
angular momenta to coincide.  There has been some effort to extend the 4D
Teukolsky formalism to the case of the Myers-Perry family of solutions.
To that end, in order to identify similar perturbative variables as in the
Teukolsky case, use of the Petrov classification for higher-dimensional
spacetimes~\cite{Coley:2004jv} has been made.  And in order to look
for decoupled equations, generalizations of the Geroch-Held-Penrose
formalism~\cite{Geroch:1973am} (a generalization of the Newman-Penrose
formalism~\cite{Newman:1961qr} for cases with algebraically-special
spacetimes) have been proposed~\cite{Durkee:2010xq}. Unfortunately,
decoupling does not occur in higher-dimensional black hole spacetimes
(see~\cite{Kunduri:2006qa,Durkee:2010qu,Murata:2011hk}), except for
near-horizon geometries (see also~\cite{Durkee:2010ea}) and some special
cases. For instance, in $D=5$ dimensions a master equation for a part
of the metric perturbations that are relevant for the study of stability
has been derived~\cite{Murata:2007gv}. For a detailed discussion of the
stability properties of black holes in $D>4$, we refer to Sec. \ref{BHs}.


Apart from the study of BH-like dark objects,
other types of dark objects have been investigated.  In particular,
this has led to the discovery of the
well-known Gregory-Laflamme instability of black strings and
p-branes~\cite{Gregory:1993vy,Gregory:1994bj}; see~\cite{Harmark:2007md}
for a review.

We also mention that there are a number of studies of Quasi-Normal
modes of black objects in 4D and in higher dimensions (for reviews
see~\cite{Cardoso:2003vt,Berti:2009kk,Konoplya:2011qq}), and that
gravitational radiation in D-dimensional spacetimes has been studied
in~\cite{Cardoso:2002pa}.

\subsection{Conclusions and Prospects for the Future}

While relativistic perturbation theory has been developed for many years
and even though
it has achieved great success in many areas of astrophysics,
cosmology, and fundamental physics, there remains a lot of work
to be done, even in the realm of 4D GR.  Among the main challenges are
the development of tools for higher-order perturbation theory, and
especially for the treatment of backreaction, which is important
for EMRIs and the particular case of cosmology where it has been invoked
to try to explain the acceleration of the Universe.  It would also be
important to extend approximation methods currently applied to GR,
to the study of
alternative theories of gravity; see Sec.~\ref{sec:alternative}.

In the case of higher-dimensional spacetimes, there has been significant
progress in the case of static non-rotating BHs; unfortunately,
such progress has not yet been extended to rotating BHs or
other black objects, where there is a pressing need for new tools.
The fact that there are no decoupled master equations that describe all
the gravitational degrees of freedom is a challenge for future work.
Given that significant progress has also been done in non-linear numerical
studies, it is important for the near future to develop
tools in perturbation theory to complement the numerical computations
and aid in their physical interpretation.
In this sense, it would be interesting to develop more sophisticated
perturbative schemes adapted to the different problems.  For instance,
the close-limit approximation could be developed for higher-dimensional
BHs in spacetimes with different asymptotic properties, which may be
of interest for physical applications like ultrarelativistic collisions
of BHs and the $AdS$/CFT conjecture in relation with quark-gluon plasmas.
Also in this line of research, point particle calculations, which model
black hole collisions when one of them is much smaller than the other,
give remarkably accurate results when extrapolated to the equal-mass case
\cite{Cardoso:2002ay,Berti:2010ce,Berti:2010gx}. Because this kind of
computation relies on linear perturbation theory, extension of classical
results to different asymptotics is clearly desirable.

\vskip 1cm